\documentclass[reprint,aps,pra,superscriptaddress,showpacs,amsmath,amssymb,floatfix]{revtex4-1}

\usepackage{txfonts}

\allowdisplaybreaks[4]

\usepackage{graphicx}

\usepackage{subfigure}
\usepackage{braket}

\def\articletitle#1{{\it #1}}

\usepackage{color}

\graphicspath{{./Figures/}}

\newcommand{\nature}{ Nature }
\newcommand{\natphoton}{ Nat.\ Photon.\ }
\newcommand{\science}{ Science }
\renewcommand{\prl}{ Phys.\ Rev.\ Lett.\ }
\newcommand{\prx}{ Phys.\ Rev.\ X }
\renewcommand{\pra}{ Phys.\ Rev.\ A }

\newcommand{\optexp}{ Opt.\ Exp.\ }

\newcommand{\repmathphys}{Rep.\ Math.\ Phys.\ }

\newcommand{\etal}{\textit{et al.}, }

\newcommand{\abs}[1]{\ensuremath{\left\lvert#1\right\rvert}}

\newcommand{\UTokyo}{
Department of Applied Physics, School of Engineering, \\
The University of Tokyo, 7-3-1 Hongo, Bunkyo-ku, Tokyo 113-8656, Japan}
\newcommand{\UTokyoQP}{
Quantum-Phase Electronics Center, School of Engineering, \\
The University of Tokyo, 7-3-1 Hongo, Bunkyo-ku, Tokyo 113-8656, Japan}
\newcommand{\UMainz}{
Institute of Physics, Staudingerweg 7,
Johannes Gutenberg-Universit\"{a}t Mainz, 55099 Mainz, Germany}
\newcommand{\ANU}{
Centre of Excellence for Quantum Computation and Communication Technology,
Department of Quantum Science, \\
Research School of Physics and Engineering,
Building 38a Science Road, \\
The Australian National University,
Acton, ACT, 2601, Australia}

\begin{document}

\title{Heralded creation of photonic qudits from parametric down conversion using linear optics}

\author{Jun-ichi Yoshikawa}
\email{yoshikawa@ap.t.u-tokyo.ac.jp}
\affiliation{\UTokyo}
\affiliation{\UTokyoQP}
\author{Marcel Bergmann}
\author{Peter van Loock}
\affiliation{\UMainz}
\author{Maria Fuwa}
\affiliation{\UTokyo}
\affiliation{\ANU}
\author{Masanori Okada}
\affiliation{\UTokyo}
\author{\\Kan Takase}
\author{Takeshi Toyama}
\author{Kenzo Makino}
\author{Shuntaro Takeda}
\author{Akira Furusawa}
\affiliation{\UTokyo}

\date{\today}

\begin{abstract}
We propose an experimental scheme to generate, in a heralded fashion, arbitrary quantum superpositions of two-mode optical states with a fixed total photon number $n$ based on weakly squeezed two-mode squeezed state resources (obtained via weak parametric down conversion), linear optics, and photon detection. 
Arbitrary $d$-level (qudit) states can be created this way where $d=n+1$. 
Furthermore, we experimentally demonstrate our scheme for $n=2$. 
The resulting qutrit states are characterized via optical homodyne tomography.
We also discuss possible extensions to more than two modes concluding that, in general, our approach ceases to work in this case. 
For illustration and with regards to possible applications, we explicitly calculate a few examples such as NOON states and logical qubit states for quantum error correction. 
In particular, our approach enables one to construct bosonic qubit error-correction codes
against amplitude damping (photon loss) with a typical suppression of $\sqrt{n}-1$ losses and spanned by two logical codewords that each correspond to an $n$-photon superposition for two bosonic modes. 
\end{abstract}


\maketitle

\section{Introduction}

Photons are an essential ingredient of most protocols for quantum information processing and quantum communication, as they can serve as carriers of ``flying quantum information'', especially in the form of flying qubits. 
However, experimentally, deterministic schemes to prepare optical quantum states remain so far within the regime of Gaussian states or classical mixtures of Gaussian states, though, in principle, third-order nonlinear optical effects or interactions with a finite-dimensional system enable one to step out of the Gaussian realm into that of non-Gaussian quantum states \cite{Yurke.prl(1986)}. 
Highly nonclassical, non-Gaussian states of traveling light, pure enough to show negative values in their Wigner functions \cite{Hudson.repmathphys(1974)}, have been created with probabilistic, heralded schemes \cite{Lvovsky.prl(2001),Lvovsky.prl(2002),Bimbard.nphoton(2010),Yukawa.oe(2013),Yukawa.pra(2013),Zavatta.prl(2006),Takeda.pra(2013),Ourjoumtsev.science(2006),Neergaard-Nielsen.prl(2006),Wakui.oe(2007)}. 
These rely on the non-Gaussianity or nonlinearity induced by a photon detection. 
Since deterministic, 100$\%$-efficient quantum nondemolition measurements of photon numbers \cite{Guerlin.nature(2007)} are currently unavailable in the optical domain, a photon detection would destroy the measured optical field. 
Nonetheless, the non-Gaussianity could still be transferred to an outgoing, propagating optical quantum state through quantum correlations.

For the optical resources before the photon detections, two-mode squeezing correlations between signal and idler fields from parametric down converters are typically utilized. 
Beyond heralding single photons \cite{Lvovsky.prl(2001)}, in previous experiments, an arbitrary superposition of photon number states up to three photons, i.e.\ $c_{0}\ket{0}+c_{1}\ket{1}+c_{2}\ket{2}+c_{3}\ket{3}$ with $c_0,c_1,c_2,c_3\in\mathbb{C}$, was experimentally generated in a heralded fashion by employing three photon detectors. Before these detections, the idler fields in the heralding lines are combined with auxiliary coherent fields \cite{Yukawa.oe(2013),Yukawa.pra(2013)}. 
On the other hand, by utilizing the interference of the idler fields in the heralding lines, arbitrary single-photon qubits encoded into two modes, i.e.\ $c_{10}\ket{10}+c_{01}\ket{01}$ with $c_{10},c_{01}\in\mathbb{C}$ (so-called dual-rail qubits), were experimentally produced \cite{Zavatta.prl(2006),Takeda.pra(2013)}.

It is then an interesting question whether we can create an arbitrary superposition of photon number states with, for instance, a total photon number of two distributed in two modes, i.e.\ $c_{20}\ket{20}+c_{11}\ket{11}+c_{02}\ket{02}$ with $c_{20},c_{11},c_{02}\in\mathbb{C}$. 
This set of quantum states forms a qutrit whose three-dimensional Hilbert space is spanned by the three basis states $\{\ket{20},\ket{11},\ket{02}\}$.
One can think of this qutrit also as a spin-1 particle with a spin value 1 corresponding to half of the total photon number $(n_1 + n_2)/2=n/2=1$ and the three possible spin projections corresponding to half of the photon number differences of the two modes, $(n_1 - n_2)/2=\{1,0,-1\}$. 
More generally, an arbitrary $d$-level spin particle can be represented by two modes with a spin value corresponding to $(n_1 + n_2)/2=n/2$ and $d=n+1$ possible spin projections corresponding to $(n_1 - n_2)/2$ (this is also referred to as the Schwinger representation).  
In the most general case, a set of number states with a total of $n$ photons distributed in $m$ modes, $\{\ket{n_1,...,n_m}\}$ with $\sum_{k=1}^mn_k=n$, spans a $d$-dimensional Hilbert space where 
\begin{align}
d = \left(\!\left(
\begin{matrix}
m\\n
\end{matrix}
\right)\!\right)
=\left(
\begin{matrix}
m+n-1\\n
\end{matrix}
\right)
=\frac{(m+n-1)!}{n!\,(m-1)!}. 
\end{align}
The qutrit above corresponds to the case of $n=2$ and $m=2$, that is one special case of the Schwinger representation with generally $m=2$ and arbitrary $n$. 
Such quantum states living in a higher-dimensional Hilbert space are important, because, for instance, a quantum error-correction code can be constructed by utilizing a certain subspace as the code space. 
For this purpose, the Hilbert space of the physical system must be big enough such that a logical quantum state can be robustly mapped between code space and error spaces.
More specifically, multiphoton states can be possibly tolerant against amplitude damping, and indeed, when qubit information is encoded, for example, as $\alpha[(\ket{40}+\ket{04})/\sqrt{2}]+\beta\ket{22}$, the information does not get lost by a random single-photon annihilation \cite{Chuang.pra(1997)}.

Here we discuss how to create such superposition states in a heralded scheme, utilizing two-mode squeezed states and linear optics in the heralding idler lines. 
As a consequence, it is shown that an arbitrary superposition with total photon number $n$ can be, in principle, created for the case of mode number $m=2$, leading to an arbitrary qudit with $d=n+1$. 
However, we also find that our scheme cannot be generally extended to the cases of $m\ge3$.
Here we also experimentally demonstrate our scheme for the two-mode qutrit case ($n=2$ and $m=2$). 
Our scheme is directly applicable to the construction of bosonic codes against amplitude damping \cite{Chuang.pra(1997)}. 
The scheme described here can be possibly combined with recent memory schemes \cite{Yoshikawa.prx(2013),Bimbard.prl(2014)}, by which the success event rate may be considerably improved. 
This paper is organized as follows. 
In Sec.~\ref{sec:herald}, we review how the creation of a heralded single photon is mathematically described, and then we discuss how this can be extended to a single-photon qubit with $n=1$ and $m=2$. 
In Sec.~\ref{sec:qutrit}, we describe the heralded creation of a qutrit for the case of $n=2$ and $m=2$, based on the factorization of a corresponding polynomial. 
In Sec.~\ref{sec:qudit}, we present a general extension of the polynomial factorization to the qudit cases of $n\ge3$ and $m=2$. 
In Sec.~\ref{sec:extension}, we discuss that further extensions of our scheme to $m\ge3$ are, in general, impossible. 
In Sec.~\ref{sec:applications}, we present a few examples and applications of our qudit generation scheme. 
In Sec.~\ref{sec:experiment}, we present an experimental demonstration of our scheme by using time bins. 
The density matrices of the heralded states are fully characterized by quantum tomography, employing homodyne detectors for the simultaneous measurements of quadrature values \cite{Takeda.pra(2013)}. 
Further examples are presented in the Appendix.

\section{Heralded Creation of an Arbitrary Qubit}
\label{sec:herald}

In a heralded scheme, typically two-mode squeezing by parametric down conversion is employed with sufficiently weak pumping, where signal and idler photons are probabilistically created in pairs. 
Mathematically, the initial two-mode squeezed state (without normalization) is expressed as $\sum_{n=0}^\infty q^n\ket{n}_s\ket{n}_i$, where weak pumping corresponds to $q\ll1$. 
Then the detection of idler photons means projection onto $\sum_{n=1}^\infty\ket{n}_i\bra{n}$.
However, in the case of very weak pumping, higher photon-number detections are unlikely, so we approximate the projection operator by $\ket{1}_i\bra{1}$. 
Alternatively, the detector can be a photon-number-resolving (PNR) detector with high efficiency, in which case the projection $\ket{1}_i\bra{1}$ is obtained even with strong pumping by discarding the cases of two or more detected idler photons. 
In addition, since we are not interested in the idler state after the measurement, in the following we express the projection measurement simply by a bra vector ${}_i\bra{1}={}_i\bra{0}\hat{a}_i$, where $\hat{a}_i$ is the annihilation operator for the idler mode. 
The process of creating a heralded single photon can then be described by
\begin{align}
& {}_i\bra{1}\sum_{n=0}^\infty q^n\ket{n}_s\ket{n}_i \notag\\
& = {}_i\bra{0}\hat{a}_i\sum_{n=0}^\infty \frac{q^n}{n!}\hat{a}_s^{\dagger n}\hat{a}_i^{\dagger n}\ket{0}_s\ket{0}_i \notag\\
& = q\hat{a}_s^\dagger\ket{0}_s = q\ket{1}_s.
\end{align}
Here, we intentionally introduced the annihilation and creation operators $\hat{a}$ and $\hat{a}^\dagger$, in prospect of later use.
Their commutation relation is $[\hat{a}_j,\hat{a}_k^\dagger]=\delta_{jk}$.
The error rate due to the approximation of $\sum_{n=1}^\infty\ket{n}_i\bra{n}$ by $\ket{1}_i\bra{1}$ based on weak pumping is of the order of $q^2$.
These higher-photon-number components turn the signal state into a mixed state.

\begin{figure}[tb]
\centering
\includegraphics{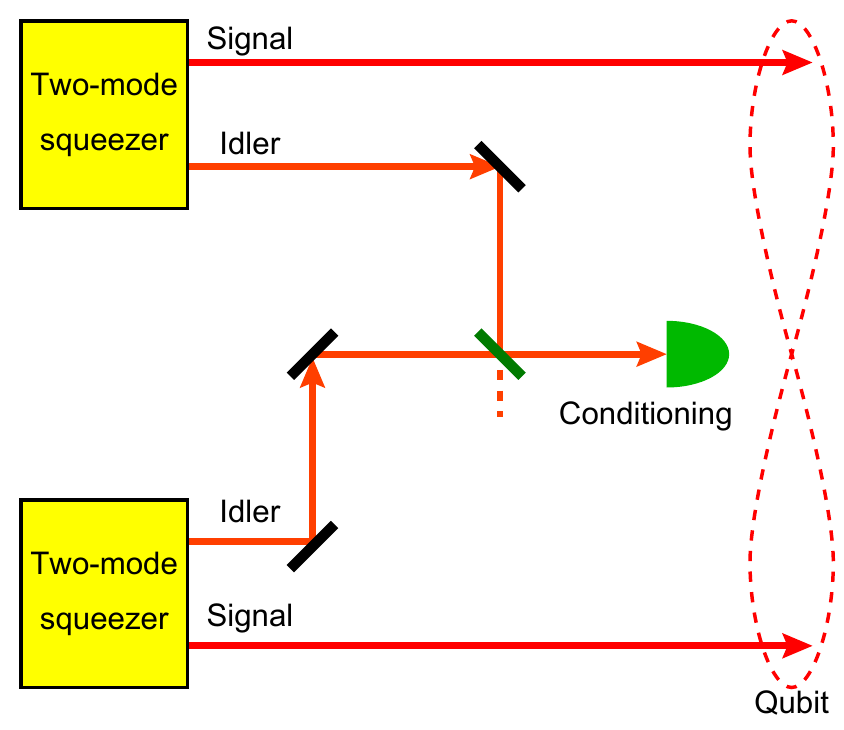}
\caption{
Scheme for creating a qubit $c_{10}\ket{10}+c_{01}\ket{01}$.
}
\label{fig:setup_qubit}
\end{figure}

An arbitrary single-photon (dual-rail) qubit $c_{10}\ket{10}+c_{01}\ket{01}$ can be created by combining two idler fields from two parametric down converters at a beam splitter before the photon detection \cite{Zavatta.prl(2006),Takeda.pra(2013)}. 
This basically means adjusting the erasure of which-path information \cite{Kim.prl(2000)}.
The scheme is illustrated in Fig.~\ref{fig:setup_qubit}. 
Note that the actual experimental demonstration was for a time-bin qubit \cite{Zavatta.prl(2006),Takeda.pra(2013)}, in which case ``two idler fields from two parametric down converters'' actually mean idler fields generated at (sufficiently) different times by a single parametric down converter. 
The scheme starts with 
\begin{align}
\sum_{n_1=0}^\infty q^{n_1}\ket{n_1}_{s1}\ket{n_1}_{i1} \otimes \sum_{n_2=0}^\infty q^{n_2}\ket{n_2}_{s2}\ket{n_2}_{i2}, 
\end{align}
and the projection by a photon detection at one output port of the beam splitter is expressed by ${}_{i1}\bra{1}{}_{i2}\bra{0}\hat{U}_{i1,i2}(t,r)$, under the approximation of neglecting the possibility of higher excitations irrelevant to the photon detection (i.e., neglecting orders $\sim q^2$). 
Alternatively, in the case of strong pumping with a PNR detector, the projection by ${}_{i2}\bra{0}$ has to be confirmed by another photon detector. 
Here, $\hat{U}_{k,\ell}(t,r)$ is a unitary operator representing a beam splitter transformation on modes $k$ and $\ell$ with a transmission coefficient $t\in\mathbb{C}$ and a reflection coefficient $r\in\mathbb{C}$, satisfying $\lvert t\rvert^2 + \lvert r\rvert^2 = 1$.
It transforms the annihilation operators as
\begin{subequations}
\begin{align}
\hat{U}_{k,\ell}^\dagger(t,r)\hat{a}_{k}\hat{U}_{k,\ell}(t,r) & = t \hat{a}_k + r \hat{a}_\ell, \\
\hat{U}_{k,\ell}^\dagger(t,r)\hat{a}_{\ell}\hat{U}_{k,\ell}(t,r) & = - r^\ast \hat{a}_k + t^\ast \hat{a}_\ell,
\end{align}
\end{subequations}
where the superscript $\ast$ denotes complex conjugate. 
Then, the projection bra vector is rewritten as 
\begin{align}
{}_{i1}\bra{1}{}_{i2}\bra{0}\hat{U}_{i1,i2}(t,r)
& = {}_{i1}\bra{0}{}_{i2}\bra{0}\hat{a}_{i1}\hat{U}_{i1,i2}(t,r) \notag\\
& = {}_{i1}\bra{0}{}_{i2}\bra{0}\hat{U}_{i1,i2}^\dagger(t,r)\hat{a}_{i1}\hat{U}_{i1,i2}(t,r) \notag\\
& = {}_{i1}\bra{0}{}_{i2}\bra{0}(t\hat{a}_{i1} + r\hat{a}_{i2}). 
\end{align}
Note that here we utilized that a two-mode vacuum state is not changed by a beam splitter, $\hat{U}_{i1,i2}(t,r)\ket{0}_{i1}\ket{0}_{i2} = \ket{0}_{i1}\ket{0}_{i2}$.
The resulting state after the projection is 
\begin{align}
& {}_{i1}\bra{1}{}_{i2}\bra{0}\hat{U}_{i1,i2}(t,r) \sum_{n_1,n_2=0}^\infty q^{n_1}q^{n_2}\ket{n_1}_{s1}\ket{n_1}_{i1}\ket{n_2}_{s2}\ket{n_2}_{i2} \notag\\
& = {}_{i1}\bra{0}{}_{i2}\bra{0}(t\hat{a}_{i1} + r\hat{a}_{i2}) \notag\\
& \quad \sum_{n_1,n_2=0}^\infty q^{n_1}q^{n_2} \frac{\hat{a}_{s1}^{\dagger n_1}\hat{a}_{i1}^{\dagger n_1}}{n_1!}\frac{\hat{a}_{s2}^{\dagger n_2}\hat{a}_{i2}^{\dagger n_2}}{n_2!} \ket{0}_{s1}\ket{0}_{i1}\ket{0}_{s2}\ket{0}_{i2} \notag\\
& = q(t \hat{a}_{s1}^\dagger + r \hat{a}_{s2}^\dagger)\ket{0}_{s1}\ket{0}_{s2} \notag\\
& = q(t \ket{1}_{s1}\ket{0}_{s2} + r \ket{0}_{s1}\ket{1}_{s2}),
\end{align}
up to the normalization factor. 
Since the coefficients $t$ and $r$ can be arbitrarily determined under the constraint $\lvert t\rvert^2 + \lvert r\rvert^2 = 1$ via the beam splitting ratio and a phase shift before the interference, an arbitrary qubit  $c_{10}\ket{1}_{s1}\ket{0}_{s2}+c_{01}\ket{0}_{s1}\ket{1}_{s2}$ is probabilistically created with this scheme. 
Because a photon detection is phase-insensitive, a phase shift after the interference is meaningless and thus only a phase shift before the interference at the beam splitter can change the argument of the complex numbers $t$ or $r$. 
For a projection onto ${}_{i1}\bra{0}{}_{i2}\bra{1}\hat{U}_{i1,i2}(t,r)$ (corresponding to the detection of one photon in the other beam splitter output port), it is easy to see that a similar calculation leads to the orthogonal dual-rail qubit state in the signal output modes, $q(-r^\ast \ket{1}_{s1}\ket{0}_{s2} + t^\ast \ket{0}_{s1}\ket{1}_{s2})$. 
Note that for the special case of $t=r=1/\sqrt{2}$, the scheme resembles the entanglement distribution in the quantum repeater protocol of Ref.~\cite{DLCZ} with the two signal modes distributed among two repeater stations.

\section{Heralded Creation of an Arbitrary Qutrit}
\label{sec:qutrit}

\begin{figure}[!tb]
\centering
\includegraphics{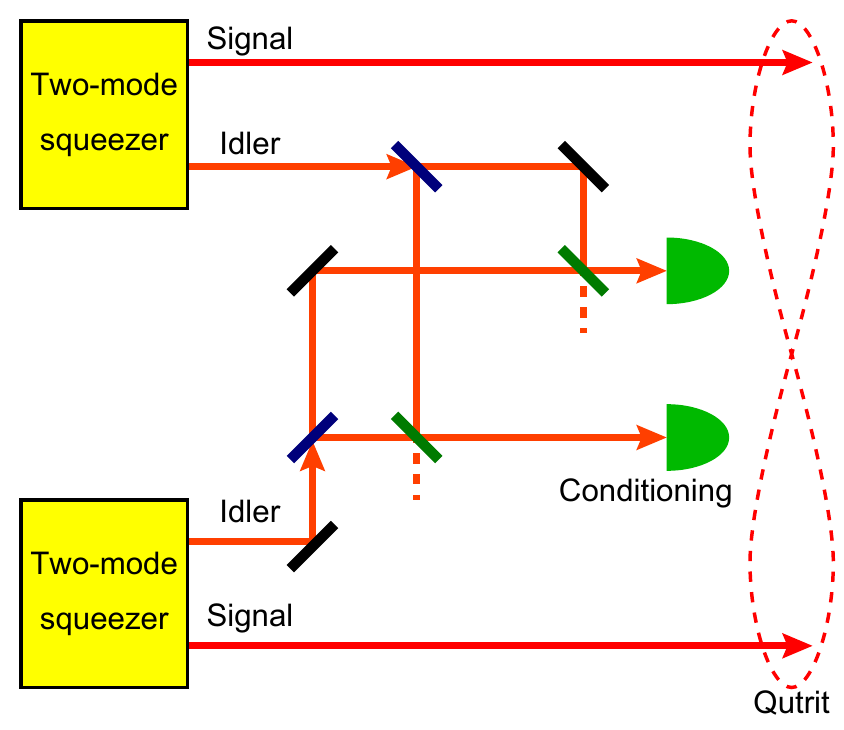}
\caption{
Scheme for creating a qutrit $c_{20}\ket{20}+c_{11}\ket{11}+c_{02}\ket{02}$.
}
\label{fig:setup_qutrit}
\end{figure}

Let us now consider the case of a qutrit with a total photon number of $n=2$ in $m=2$ modes, i.e.,
\begin{align}
& c_{20}\ket{20}_{1,2}+c_{11}\ket{11}_{1,2}+c_{02}\ket{02}_{1,2} \notag\\
& = \biggl(\frac{c_{20}}{\sqrt{2}}\hat{a}_1^{\dagger2}+c_{11}\hat{a}_1^\dagger\hat{a}_2^\dagger+\frac{c_{02}}{\sqrt{2}}\hat{a}_2^{\dagger2}\biggr)\ket{00}_{1,2}.
\end{align}
The crucial observation is the following decomposition,
\begin{align}
& c_{20}\hat{a}_1^{\dagger2}+\sqrt{2}c_{11}\hat{a}_1^\dagger\hat{a}_2^\dagger+c_{02}\hat{a}_2^{\dagger2} \notag\\
& = (d_{11}\hat{a}_1^\dagger+d_{12}\hat{a}_2^\dagger)(d_{21}\hat{a}_1^\dagger+d_{22}\hat{a}_2^\dagger),
\end{align}
where $c_{20},c_{11},c_{02}\in\mathbb{C}$, $d_{11},d_{12},d_{21},d_{22}\in\mathbb{C}$.
Since all the creation operators commute with each other, this decomposition corresponds to the problem of factorizing polynomials with complex coefficients,
\begin{align}
c_{20}z^2+\sqrt{2}c_{11}z+c_{02}=(d_{11}z+d_{12})(d_{21}z+d_{22}).
\end{align}
However, this factorization is always possible from the fundamental theorem of algebra.
The answer of $az^2+bz+c=0$ is $z=(-b\pm\sqrt{b^2-4ac})/2a$ when $a\neq0$, where $\sqrt{|u|e^{i\theta}}$ is either $\sqrt{|u|}e^{i\theta/2}$ or $\sqrt{|u|}e^{i\pi+i\theta/2}$.

For implementing the product of two 1st-order terms $d_{11}\hat{a}_1^\dagger+d_{12}\hat{a}_2^\dagger$ and $d_{21}\hat{a}_1^\dagger+d_{22}\hat{a}_2^\dagger$, the simplest way is to split each of the two idler modes into two by a beam splitter. 
The scheme is depicted in Fig.~\ref{fig:setup_qutrit}. 
We introduce two ancilla modes in a vacuum state $\ket{0}_{a1}\ket{0}_{a2}$ which enter the unused ports of the beam splitters. 
Then we combine the two split idler modes at two beam splitters with different transmission and reflection coefficients $(t_1,r_1)$ and $(t_2,r_2)$. 
More specifically, here we take the beam splitter operation as
\begin{align}
& \hat{U}_{i1,i2}(t_1,r_1)\hat{U}_{a1,a2}(t_2,r_2) \notag\\
& \quad \times\hat{U}_{i1,a1}\biggl(\frac{1}{\sqrt{2}},-\frac{1}{\sqrt{2}}\biggr)\hat{U}_{i2,a2}\biggl(\frac{1}{\sqrt{2}},-\frac{1}{\sqrt{2}}\biggr),
\end{align}
followed by the projection ${}_{i1,i2,a1,a2}\bra{1,0,1,0}$.

Eventually, the two-mode signal state after the projection becomes 
\begin{widetext}
\begin{align}
& {}_{i1,i2,a1,a2}\bra{1,0,1,0}\hat{U}_{i1,i2}(t_1,r_1)\hat{U}_{a1,a2}(t_2,r_2)\hat{U}_{i1,a1}\biggl(\frac{1}{\sqrt{2}},-\frac{1}{\sqrt{2}}\biggr)\hat{U}_{i2,a2}\biggl(\frac{1}{\sqrt{2}},-\frac{1}{\sqrt{2}}\biggr) 
\sum_{n_1,n_2=0}^\infty q^{n_1}q^{n_2}\ket{n_1,n_1,n_2,n_2}_{s1,i1,s2,i2}\ket{0,0}_{a1,a2} \notag\\
& = {}_{i1,i2,a1,a2}\bra{0,0,0,0}(t_1\hat{a}_{i1}+r_1\hat{a}_{i2})(t_2\hat{a}_{a1}+r_2\hat{a}_{a2})\hat{U}_{i1,a1}\biggl(\frac{1}{\sqrt{2}},-\frac{1}{\sqrt{2}}\biggr)\hat{U}_{i2,a2}\biggl(\frac{1}{\sqrt{2}},-\frac{1}{\sqrt{2}}\biggr) \notag\\
& \qquad\qquad
\sum_{n_1,n_2=0}^\infty q^{n_1}q^{n_2}\ket{n_1,n_1,n_2,n_2}_{s1,i1,s2,i2}\ket{0,0}_{a1,a2} \notag\\
& = {}_{i1,i2,a1,a2}\bra{0,0,0,0}\biggl[t_1\biggl(\frac{\hat{a}_{i1}-\hat{a}_{a1}}{\sqrt{2}}\biggr)+r_1\biggl(\frac{\hat{a}_{i2}-\hat{a}_{a2}}{\sqrt{2}}\biggr)\biggr]\biggl[t_2\biggl(\frac{\hat{a}_{i1}+\hat{a}_{a1}}{\sqrt{2}}\biggr)+r_2\biggl(\frac{\hat{a}_{i2}+\hat{a}_{a2}}{\sqrt{2}}\biggr)\biggr] \notag\\
& \qquad\qquad \sum_{n_1,n_2=0}^\infty q^{n_1}q^{n_2}\frac{\hat{a}_{s1}^{\dagger n_1}\hat{a}_{i1}^{\dagger n_1}}{n_1!}\frac{\hat{a}_{s2}^{\dagger n_2}\hat{a}_{i2}^{\dagger n_2}}{n_2!}\ket{0,0,0,0,0,0}_{s1,i1,s2,i2,a1,a2} \notag\\
& = \frac{q^2}{2}(t_1t_2\hat{a}_{s1}^{\dagger2}+t_1r_2\hat{a}_{s1}^\dagger\hat{a}_{s2}^\dagger+r_1t_2\hat{a}_{s2}^\dagger\hat{a}_{s1}^\dagger+r_1r_2\hat{a}_{s2}^{\dagger2})\ket{0,0}_{s1,s2} \notag\\
& = \frac{q^2}{2}(t_1\hat{a}_{s1}^\dagger+r_1\hat{a}_{s2}^\dagger)(t_2\hat{a}_{s1}^\dagger+r_2\hat{a}_{s2}^\dagger)\ket{0,0}_{s1,s2}.
\label{eq:qutrit}
\end{align}
\end{widetext}
Note that $\bra{0}\hat{a}^2\hat{a}^{\dagger2}\ket{0}=2$ is used in the above calculation.

\subsection{Interpretation of the bias factor}
\label{ssec:bias}

The first excitation $\hat{b}_2^\dagger\coloneqq t_2\hat{a}_1^\dagger+r_2\hat{a}_2^\dagger$ can be decomposed into a part parallel to the second excitation $\hat{b}_1^\dagger\coloneqq t_1\hat{a}_1^\dagger+r_1\hat{a}_2^\dagger$ and an orthogonal part $\hat{b}_{1\perp}^\dagger\coloneqq -r_1\hat{a}_1^\dagger+t_1\hat{a}_2^\dagger$ as
\begin{align}
\hat{b}_2^\dagger = c_\parallel\hat{b}_1^\dagger+c_\perp\hat{b}_{1\perp}^\dagger.
\end{align}
However, the parallel part has a twice as large contribution as the orthogonal part,
\begin{align}
&\hat{b}_1^\dagger\hat{b}_2^\dagger\ket{0,0}_{\hat{b}_1,\hat{b}_{1\perp}} \notag\\
& = \hat{b}_1^\dagger(c_\parallel\hat{b}_1^\dagger+c_\perp\hat{b}_{1\perp}^\dagger)\ket{0,0}_{\hat{b}_1,\hat{b}_{1\perp}} \notag\\
& = \sqrt{2}c_\parallel\ket{2,0}_{\hat{b}_1,\hat{b}_{1\perp}}+c_\perp\ket{1,1}_{\hat{b}_1,\hat{b}_{1\perp}},
\end{align}
where the two-mode representation is appropriately chosen so that $\hat{b}_1^\dagger\ket{0,0}_{\hat{b}_1,\hat{b}_{1\perp}}=\ket{1,0}_{\hat{b}_1,\hat{b}_{1\perp}}$ and $\hat{b}_{1\perp}^\dagger\ket{0,0}_{\hat{b}_1,\hat{b}_{1\perp}}=\ket{0,1}_{\hat{b}_1,\hat{b}_{1\perp}}$.

This ``bias'' is explained as follows.
When $\hat{b}_2^\dagger$ is orthogonal to $\hat{b}_1^\dagger$, the photon that heralds $\hat{b}_1^\dagger$ must, after the beam splitter network, go to the detector ${}_{i1}\bra{1}$, while the photon that heralds $\hat{b}_2^\dagger$ must go to the detector ${}_{a1}\bra{1}$.
On the other hand, when $\hat{b}_2^\dagger$ is parallel to $\hat{b}_1^\dagger$, the photon that heralds $\hat{b}_1^\dagger$ may go to either of ${}_{i1}\bra{1}$ and ${}_{a1}\bra{1}$, while the photon that heralds $\hat{b}_2^\dagger$ must go to the other detector.
This freedom of swapping photons increases the contribution of parallel components.
Although the initial two-mode squeezed state $\sum_{n_1,n_2=0}^\infty q^{n_1}q^{n_2}\ket{n_1,n_1,n_2,n_2}_{s1,i1,s2,i2}$ equally contains all the two-photon signal state $c_{2,0}\ket{2,0}_{s1,s2}+c_{1,1}\ket{1,1}_{s1,s2}+c_{0,2}\ket{0,2}_{s1,s2}$ with the probability density $O(q^4)$,
$\hat{b}_1^{\dagger2}$ is twice more likely to be heralded than $\hat{b}_1^\dagger\hat{b}_{1\perp}^\dagger$.

This observation is naturally extended to the case of general total photon numbers $n$, which is described in Sec.~\ref{sec:qudit}, in which case $\hat{b}_1^{\dagger n-k}\hat{b}_{1\perp}^{\dagger k}$ has a contribution proportional to $(n-k)!k!$.

\section{Heralded Creation of an Arbitrary Qudit}
\label{sec:qudit}

The fundamental theorem of algebra says that an arbitrary non-constant single-variable polynomial with complex coefficients has at least one complex root.
From this, it can be derived that the decomposition into 1st-order terms,
\begin{align}
& c_nz^n+c_{n-1}z^{n-1}+...+c_1z+c_0 \notag\\
& = c_n(z+d_1)(z+d_2)...(z+d_n),
\end{align}
is always possible when $c_n\neq0$.

Therefore, an arbitrary superposition state can be, in principle, decomposed as a product of 1st-order creation terms on a vacuum state,
\begin{align}
& \sum_{k=0}^n c_{(n-k)k}\ket{n-k}_1\ket{k}_2 \notag\\
& = \sum_{k=0}^n \frac{c_{(n-k)k}}{\sqrt{(n-k)!k!}} \hat{a}_1^{\dagger n-k}\hat{a}_2^{\dagger k}\ket{0}_1\ket{0}_2 \notag\\
& = \biggl[\prod_{k=0}^n (d_{k,1}\hat{a}_1^\dagger+d_{k,2}\hat{a}_2^\dagger)\biggr]\,\ket{0}_1\ket{0}_2.
\end{align}

This is created via a heralded scheme that is a natural extension of the two-photon qutrit case in Sec.~\ref{sec:qutrit}.
We first equally split each of the two idler modes into $n$ modes by a series of beam splitters.
Mathematically, for each idler mode $k = 1,2$, we introduce $n-1$ ancilla vacuum modes, and we combine them at a series of beam splitters, which is, for instance, described by
\begin{align}
\hat{\mathcal{U}}_k =
& \hat{U}_{ik,ak1}\biggl(\frac{1}{\sqrt{2}},-\frac{1}{\sqrt{2}}\biggr)\hat{U}_{ik,ak2}\biggl(\sqrt{\frac{2}{3}},-\frac{1}{\sqrt{3}}\biggr) \notag\\
&\quad ... \hat{U}_{ik,ak(n-1)}\biggl(\sqrt{\frac{n-1}{n}},-\frac{1}{\sqrt{n}}\biggr).
\end{align}
This network divides the idler modes with an equal weight,
\begin{align}
[\hat{\mathcal{U}}_k^\dagger\hat{a}_{ik}\hat{\mathcal{U}}_k,\hat{a}_{ik}^\dagger] = [\hat{\mathcal{U}}_k^\dagger\hat{a}_{ak\ell}\hat{\mathcal{U}}_k,\hat{a}_{ik}^\dagger] = \frac{1}{\sqrt{n}},
\end{align}
for $\ell = 1,...,n-1$.
Then we combine the two sets of the split idler modes at $n$ beam splitters $\hat{U}_{i1,i2}(t_1,r_1)$, $\hat{U}_{a11,a21}(t_2,r_2)$, ..., $\hat{U}_{a1(n-1),a2(n-1)}(t_n,r_n)$, each followed by a photon detection.
The whole procedure can be expressed as
\begin{widetext}
\begin{align}
& {}_{i1,i2,a11,a21,...,a1(n-1),a2(n-1)}\bra{1,0,1,0,...,1,0}\hat{U}_{i1,i2}(t_1,r_1)\hat{U}_{a11,a21}(t_2,r_2)...\hat{U}_{a1(n-1),a2(n-1)}(t_n,r_n)\hat{\mathcal{U}}_1\hat{\mathcal{U}}_2 \notag\\
&\qquad\qquad \sum_{n_1,n_2=0}^\infty q^{n_1}q^{n_2}\ket{n_1,n_1,n_2,n_2}_{s1,i1,s2,i2}\ket{0,0,...,0,0}_{a11,a21,...,a1(n-1),a2(n-1)} \notag\\
&\quad = {}_{i1,i2,a11,a21,...,a1(n-1),a2(n-1)}\bra{0,0,0,0,\dots,0,0}
(t_1\hat{\mathcal{U}}_1^\dagger\hat{a}_{i1}\hat{\mathcal{U}}_1+r_1\hat{\mathcal{U}}_2^\dagger\hat{a}_{i2}\hat{\mathcal{U}}_2)
(t_2\hat{\mathcal{U}}_1^\dagger\hat{a}_{a11}\hat{\mathcal{U}}_1+r_2\hat{\mathcal{U}}_2^\dagger\hat{a}_{a21}\hat{\mathcal{U}}_2)... \notag\\
&\qquad\qquad (t_n\hat{\mathcal{U}}_1^\dagger\hat{a}_{a1(n-1)}\hat{\mathcal{U}}_1+r_n\hat{\mathcal{U}}_2^\dagger\hat{a}_{a2(n-1)}\hat{\mathcal{U}}_2)
\sum_{n_1,n_2=0}^\infty q^{n_1}q^{n_2}\frac{\hat{a}_{s1}^{\dagger n_1}\hat{a}_{i1}^{\dagger n_1}}{n_1!}\frac{\hat{a}_{s2}^{\dagger n_2}\hat{a}_{i2}^{\dagger n_2}}{n_2!}\ket{0,0,0,0,0,0,...,0,0}_{s1,i1,s2,i2,a11,a21,...,a1(n-1),a2(n-1)} \notag\\
&\quad = \frac{q^n}{n^{n/2}}(t_1\hat{a}_{s1}^\dagger+r_1\hat{a}_{s2}^\dagger)...(t_n\hat{a}_{s1}^\dagger+r_n\hat{a}_{s2}^\dagger)\ket{0,0}_{s1,s2}.
\end{align}
\end{widetext}
Note that $\bra{0}\hat{a}^n\hat{a}^{\dagger n}\ket{0}=n!$ is used in the above calculation.

Like above, the heralded creation of a qudit with $n$ photons is, in principle, possible based on the decomposition into 1st-order terms. However, finding the set of decomposition coefficients $\{d_{k\ell}\}$ for a specific case $\{c_{k\ell}\}$ is not an easy problem in general, as well as the factorization of the $n$-th order polynomial.

\begin{figure}[tb]
\centering
\includegraphics{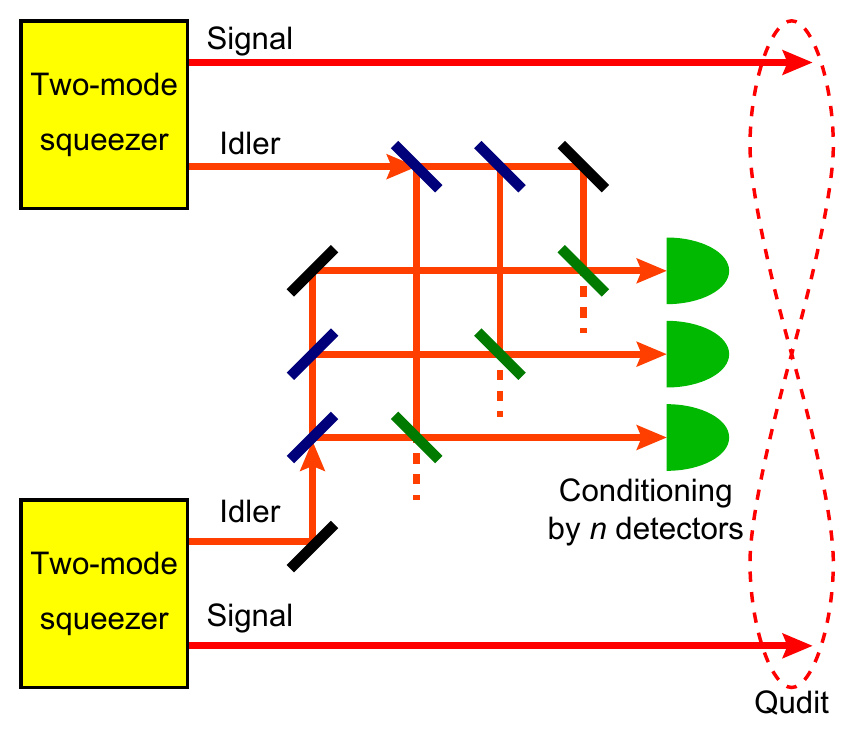}
\caption{
Scheme for creating a qudit $\sum_{k=0}^n c_{(n-k)k}\ket{n-k}\ket{k}$, where the dimension is $d=n+1$.
The number of the conditioning detectors corresponds to the total photon number $n$.
The figure, as an example, is for $n=3$.
}
\label{fig:setup_qudit}
\end{figure}

\section{Possibility of further multi-mode extensions}
\label{sec:extension}

So far, we have discussed that an arbitrary superposition state with an arbitrary total photon number $n$ can be, in principle, created with heralded schemes when the number of modes is $m=2$. 
Similarly, we may consider an extension to a general number of modes $m$ as 
\begin{align}
\sum_{n_1+...+n_m = n}c_{n_1,...,n_m}\ket{n_1,...,n_m}_{1...m}. 
\end{align}
In this general case, we have to consider the factorization of a polynomial with $m-1$  variables, 
\begin{align}
\sum_{n_1+...+n_m = n}\frac{c_{n_1,...,n_m}}{\sqrt{n_1!...n_m!}}z_1^{n_1}...z_{m-1}^{n_{m-1}}. 
\label{eq:poly_totalNArb_MArb}
\end{align}
In the case where a factorization into 1st-order terms $(\sum_{k=1}^{m-1}d_kz_k)+d_m$ is possible, the corresponding state can be again
created in a similar manner by utilizing beam splitter networks before heralding photon detections.

However, it may not be possible to factorize polynomials when they contain more than two variables. 
Therefore, the cases with more than three modes $m\ge3$ is an open question except for the trivial case of a total photon number $n\le1$. 
In fact, the insufficient degrees of freedom imply the requirement of a totally different scheme instead of the factorization into 1st-order terms. 
For instance, for the simplest case of $n=2$ and $m=3$, the polynomial to be factorized (absorbing the factors of $1/\sqrt{n_1!n_2!n_3!}$ into the coefficients $c_{n_1,n_2,n_3}$) is 
\begin{align}
c_{2,0,0}z_1^2+c_{1,1,0}z_1z_2+c_{0,2,0}z_2^2+c_{1,0,1}z_1+c_{0,1,1}z_2+c_{0,0,2},
\end{align}
while that after the factorization is
\begin{align}
c_{0,0,2}(1+d_{1,1}z_1+d_{1,2}z_2)(1+d_{2,1}z_1+d_{2,2}z_2).
\end{align}
Obviously, the degree of freedom after the factorization is not sufficient to cover all the 2nd-order polynomials.

Similarly, we can consider the extension to an arbitrary superposition state containing no more than $n$ photons in total, distributed in $m$ modes,
\begin{align}
\sum_{n_1+...+n_m \le n}c_{n_1,...,n_m}\ket{n_1,...,n_m}_{1,...,m}. 
\end{align}
In this case, the desired decomposition is
\begin{align}
& \sum_{n_1+...+n_m \le n} \frac{c_{n_1,...,n_m}}{\sqrt{n_1...n_mk!}} \hat{a}_1^{\dagger n_1}...\hat{a}_m^{\dagger n_m} \notag\\
& = \prod_{k=0}^n (d_{k,1}\hat{a}_1^\dagger+...+d_{k,m}\hat{a}_m^\dagger+d_{k,m+1}). 
\end{align}
If the above decomposition is possible, each of the single excitations $d_{k,1}\hat{a}_1^\dagger+...+d_{k,m}\hat{a}_m^\dagger+d_{k,m+1}$ is, in principle, possible, since addition of a zeroth-order term $d_{k,m+1}$ to first-order terms $d_{k,1}\hat{a}_1^\dagger+...+d_{k,m}\hat{a}_m^\dagger$ is possible by a small coherent displacement of the corresponding idler mode in phase space $\hat{a}_i \to \hat{a}_i+\epsilon_i$ before the photon detection \cite{Lvovsky.prl(2002),Bimbard.nphoton(2010),Yukawa.oe(2013)}.
The corresponding polynomial to be factorized becomes
\begin{align}
\sum_{n_1+...+n_m \le n}\frac{c_{n_1,...,n_m}}{\sqrt{n_1!...n_m!}}z_1^{n_1}...z_m^{n_1}. 
\label{eq:poly_maxNArb_MArb}
\end{align}
The polynomial of Eq.~\eqref{eq:poly_totalNArb_MArb} is equivalent to that of Eq.~\eqref{eq:poly_maxNArb_MArb} when $m$ is replaced by $m+1$, and thus the problem of ``up to $n$ photons in $m$ modes'' is equivalent to that of ``total $n$ photons in $m+1$ modes''.

\section{Examples and Applications}
\label{sec:applications}

\subsection{Error-correction code for loss}

An important possible application of our general superposition states is the creation of quantum error-correction codewords and their logical states. 
Taking advantage of our scheme to prepare multiphoton states, here we consider an error-correction code against amplitude damping. 
A famous example is a logical qubit defined as 
\begin{subequations}
\begin{align}
\ket{0}_L &=\frac{1}{\sqrt{2}}\left( \ket{40} + \ket{04}\right), \\
\ket{1}_L &= \ket{22}.
\label{eq:4photonlosscode}
\end{align}
\end{subequations}

Then, $\ket{\Psi} = \alpha\ket{0}_L+\beta\ket{1}_L$ is a good encoding of a qubit against a random one-photon loss \cite{Chuang.pra(1997)}. 
Assuming $\alpha,\beta \neq 0$, the logical qubit state can be expressed in terms of creation operators, 
\begin{align}
& \frac{\alpha}{\sqrt{2}}(\ket{40}+\ket{04})+\beta \ket{22} \notag\\
& = \left[ \frac{\alpha}{\sqrt{2}} \left(\frac{a_{1}^{\dagger 4}}{\sqrt{4!}} + \frac{a_{2}^{\dagger 4}}{\sqrt{4!}} \right) + \frac{\beta}{2}a_{1}^{\dagger 2}a_{2}^{\dagger 2}\right]\ket{00} \notag\\
& = \left(\frac{\alpha}{4\sqrt{3}}a_{1}^{\dagger 4}+\frac{\alpha}{4\sqrt{3}}a_{2}^{\dagger 4}+ \frac{\beta}{2}a_{1}^{\dagger 2}a_{2}^{\dagger 2}\right)\ket{00} \notag\\
& \eqqcolon p(a_{1}^{\dagger},a_{2}^{\dagger})\ket{00}.
\end{align}

\begin{figure*}[tb]
\centering
\subfigure[~$N=2$]{\includegraphics[scale=0.3]{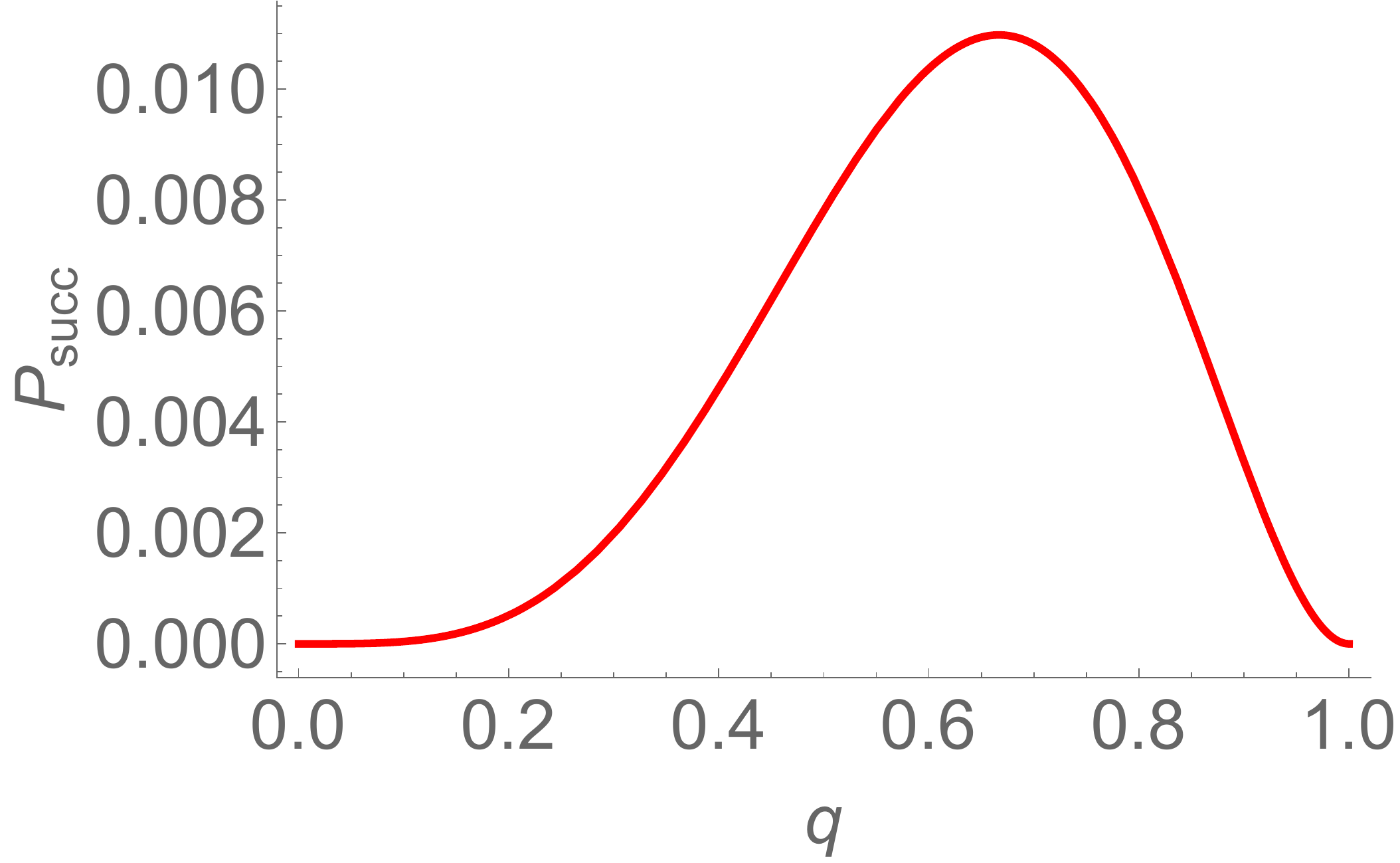}}\quad
\subfigure[~$N=3$ (orange), $N=4$ (blue), $N=5$ (green) ]{\includegraphics[scale=0.3]{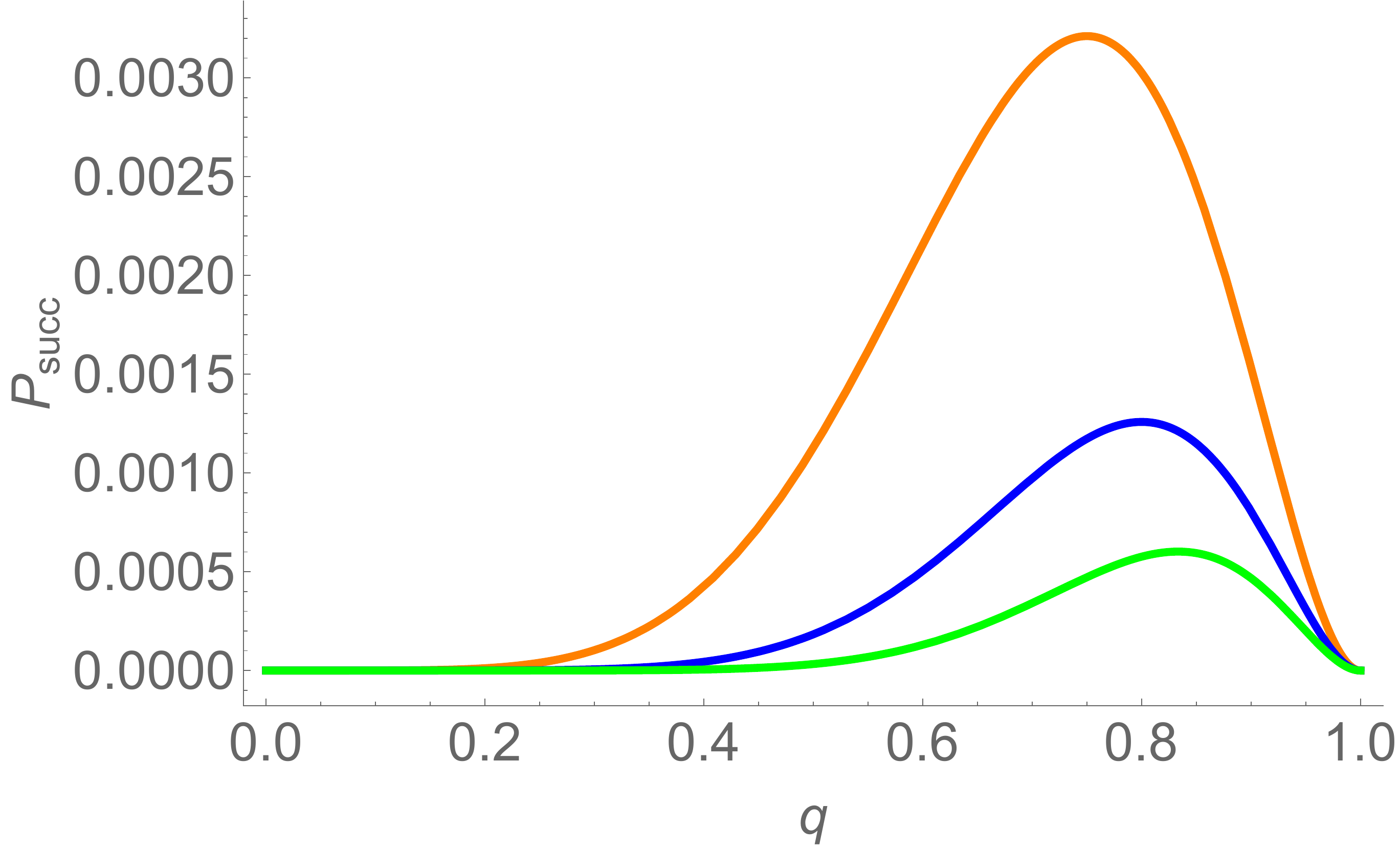}}\quad
\caption{
Success probability for creating $\frac{1}{\sqrt{2}}(\ket{N0}+\ket{0N})$ for various $N$ in dependance of $q$ (PNRD).
Note that the $N=2$ NOON state can be also directly obtained from two single-photon states using a beam splitter.}
\label{fig:NOON}
\end{figure*}

To find the transmittance and the reflection coefficients in Eq.~\eqref{eq:qutrit}, one has to determine the decomposition of $p(a_{1}^{\dagger},a_{2}^{\dagger})$ into linear factors. 
A short calculation shows
\begin{widetext}
\begin{align}
p(a_{1}^{\dagger},a_{2}^{\dagger})\ket{00}
= & \frac{\alpha}{4\sqrt{3}}\left(a_{1}^{\dagger}- a_{2}^{\dagger}\sqrt{-\frac{\sqrt{3}\beta}{\alpha}+ \sqrt{\frac{3\beta^{2}}{\alpha^{2}}-1}}\right)
\left(a_{1}^{\dagger}- a_{2}^{\dagger}\sqrt{-\frac{\sqrt{3}\beta}{\alpha}- \sqrt{\frac{3\beta^{2}}{\alpha^{2}}-1}}\right) \notag\\
& \times \left(a_{1}^{\dagger}+ a_{2}^{\dagger}\sqrt{-\frac{\sqrt{3}\beta}{\alpha} + \sqrt{\frac{3\beta^{2}}{\alpha^{2}}-1}}\right) 
\left(a_{1}^{\dagger}+a_{2}^{\dagger}\sqrt{-\frac{\sqrt{3}\beta}{\alpha} - \sqrt{\frac{3\beta^{2}}{\alpha^{2}}-1}}\right)
\ket{00}.
\end{align}

The expression is not yet in the form of Eq.~\eqref{eq:qutrit}.
This is done by rescaling each linear factor to obtain the transmission and reflection coefficients:
\begin{subequations}
\begin{align}
t_{1} = t_{3}
= & \frac{1}{\sqrt{1+\abs{-\frac{\sqrt{3}\beta}{\alpha}+\sqrt{\frac{3\beta^{2}}{\alpha^{2}}-1}}^{2}}}, &
r_{1} = r_{3}
= & \frac{-\frac{\sqrt{3}\beta}{\alpha}+\sqrt{\frac{3\beta^{2}}{\alpha^{2}}-1}}{\sqrt{1+\abs{-\frac{\sqrt{3}\beta}{\alpha}+\sqrt{\frac{3\beta^{2}}{\alpha^{2}}-1}}^{2}}}, \\
t_{2} = t_{4}
= & \frac{1}{\sqrt{1+\abs{\frac{\sqrt{3}\beta}{\alpha}+\sqrt{\frac{3\beta^{2}}{\alpha^{2}}-1}}^{2}}}, &
r_{2} = r_{4}
= & \frac{-\frac{\sqrt{3}\beta}{\alpha}-\sqrt{\frac{3\beta^{2}}{\alpha^{2}}-1}}{\sqrt{1+\abs{\frac{\sqrt{3}\beta}{\alpha}+\sqrt{\frac{3\beta^{2}}{\alpha^{2}}-1}}^{2}}}.
\end{align}
\end{subequations}
The success probability for obtaining the desired heralded state is found to be
\begin{equation}
P_\text{succ}(\alpha,\beta)=\frac{48}{\alpha^{2}}\frac{q^{8}}{256}(1-q^{2})^{2}\left(1+\abs{-\frac{\sqrt{3}\beta}{\alpha}+\sqrt{\frac{3\beta^{2}}{\alpha^{2}}-1}}^{2}\right)^{-1}
\left(1+\abs{\frac{\sqrt{3}\beta}{\alpha}+\sqrt{\frac{3\beta^{2}}{\alpha^{2}}-1}}^{2}\right)^{-1}.
\end{equation}
\end{widetext}

\begin{figure}[tb]
\centering
\includegraphics[scale=0.6]{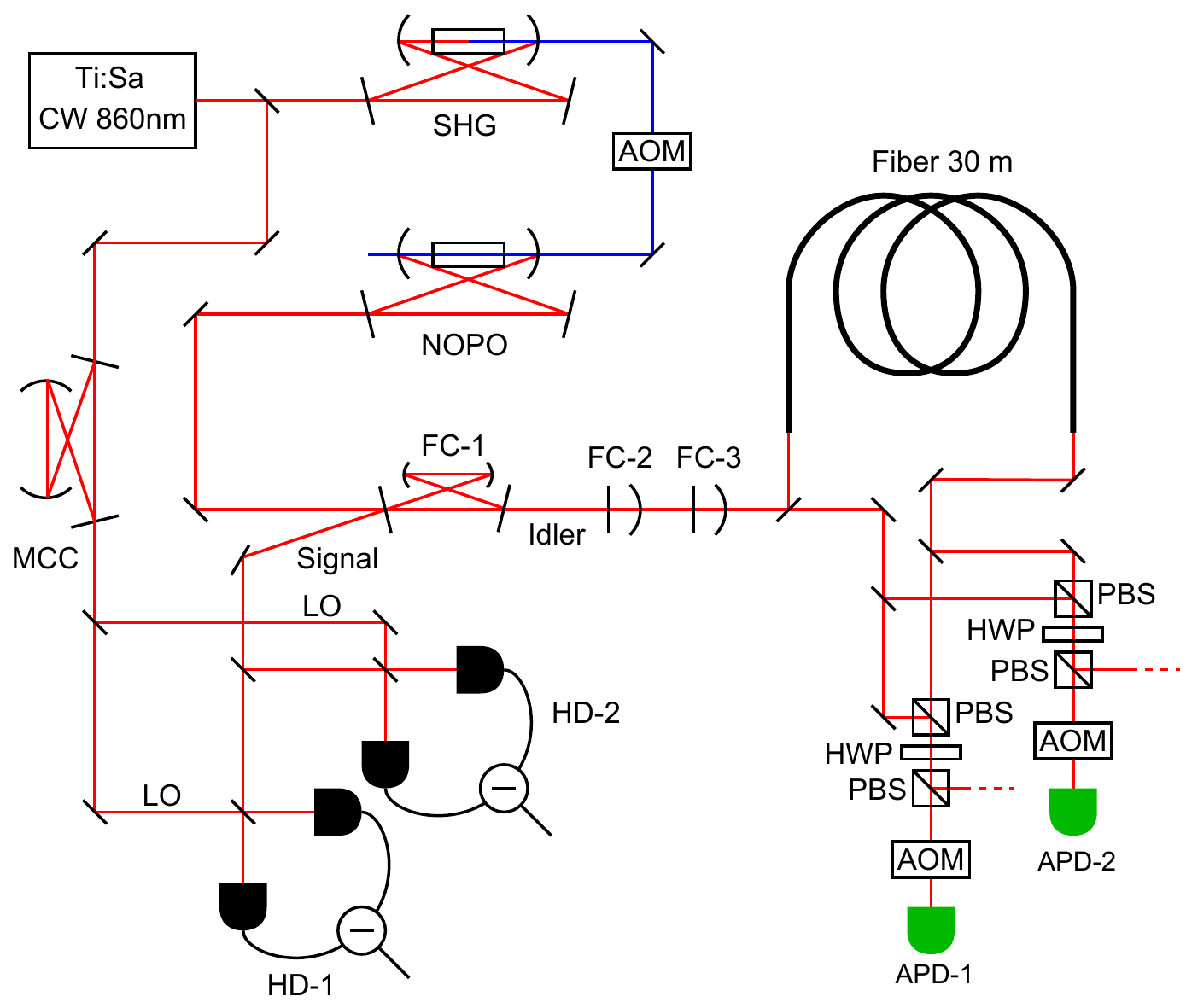}
\caption{
Experimental setup. Ti:Sa denotes titanium sapphire laser, CW continuous-wave, SHG second harmonic generator, AOM acousto-optic modulator, FC filter cavity, MCC spatial-mode cleaning cavity, LO local oscillator, and HD homodyne detector.
}
\label{fig:experiment}
\end{figure}
\begin{figure*}[tb]
\centering
\subfigure[Density matrix (real).]{\includegraphics[scale=0.4, clip]{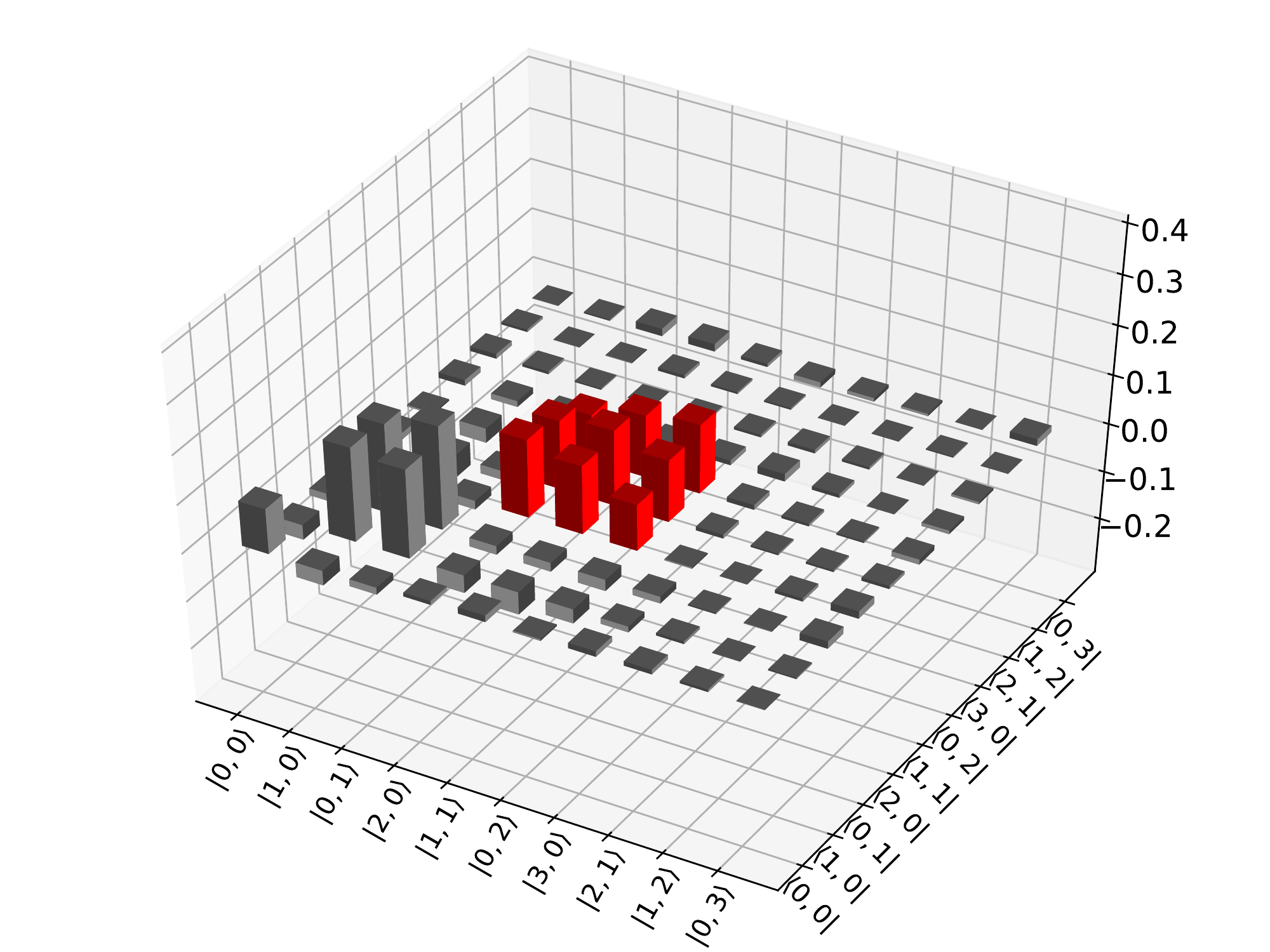}}
\subfigure[Density matrix (imaginary).]{\includegraphics[scale=0.4, clip]{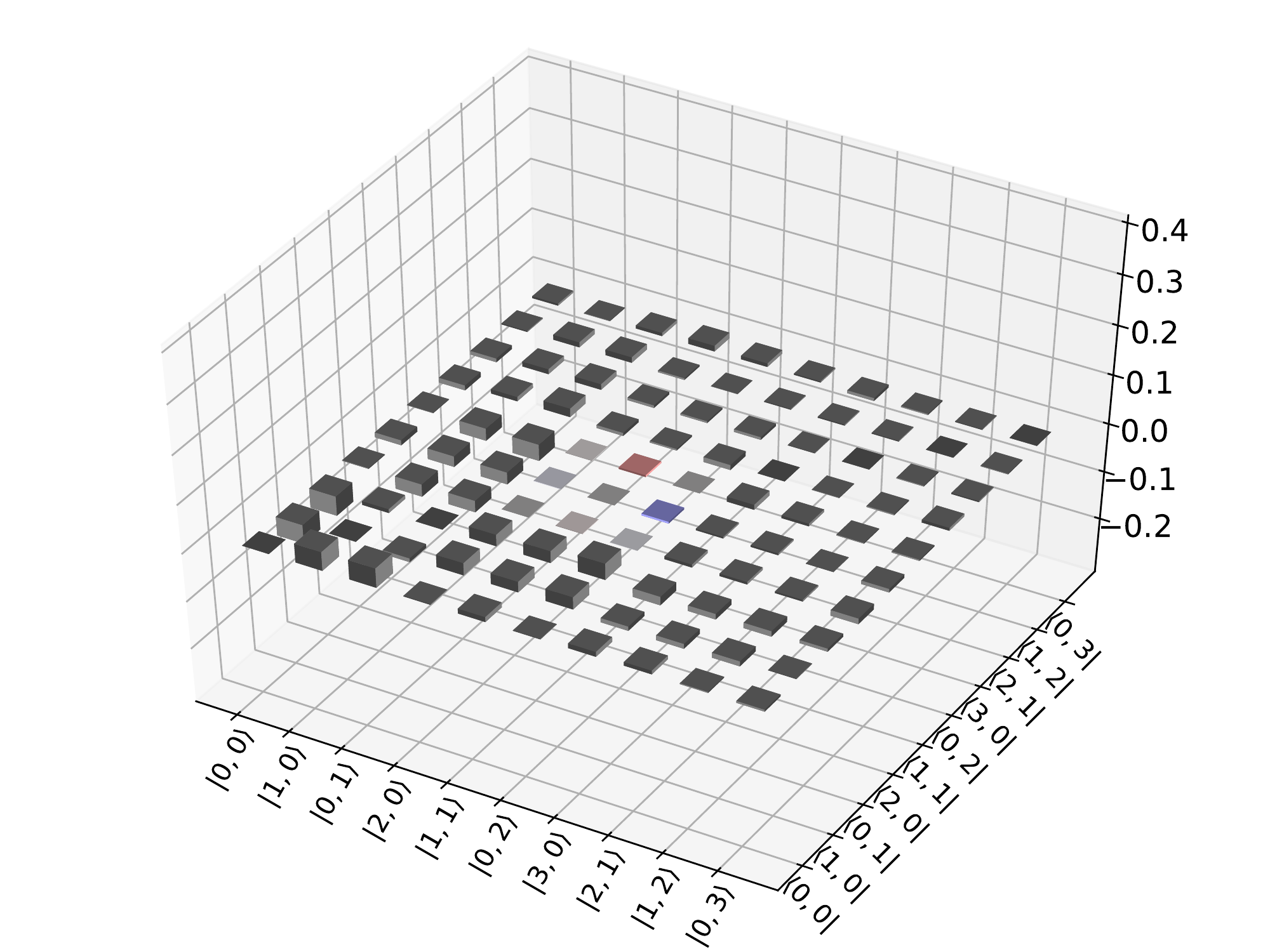}}
\subfigure[Qutrit subspace (real).]{\includegraphics[scale=0.35, clip]{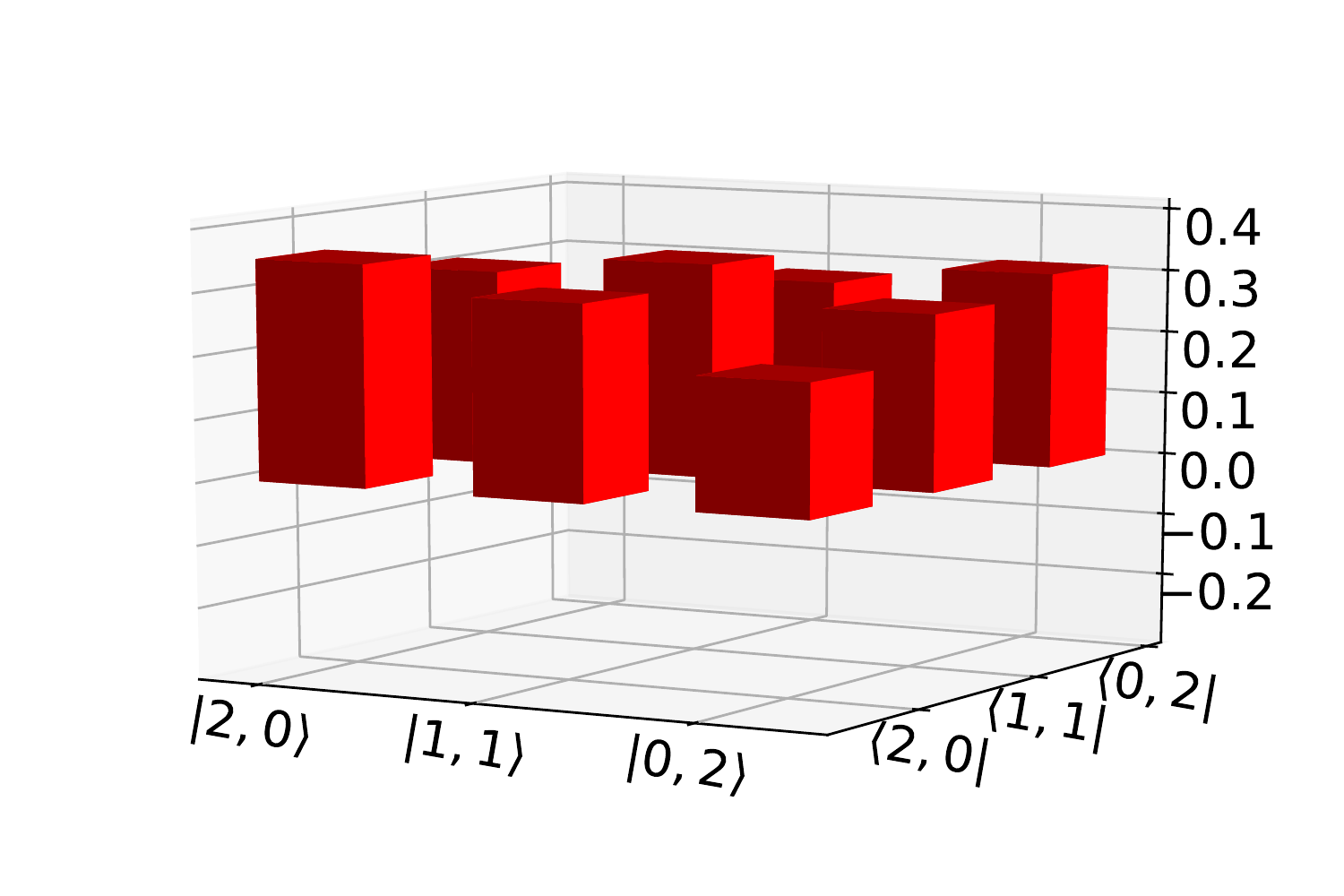}}
\subfigure[Qutrit subspace (imaginary).]{\includegraphics[scale=0.35, clip]{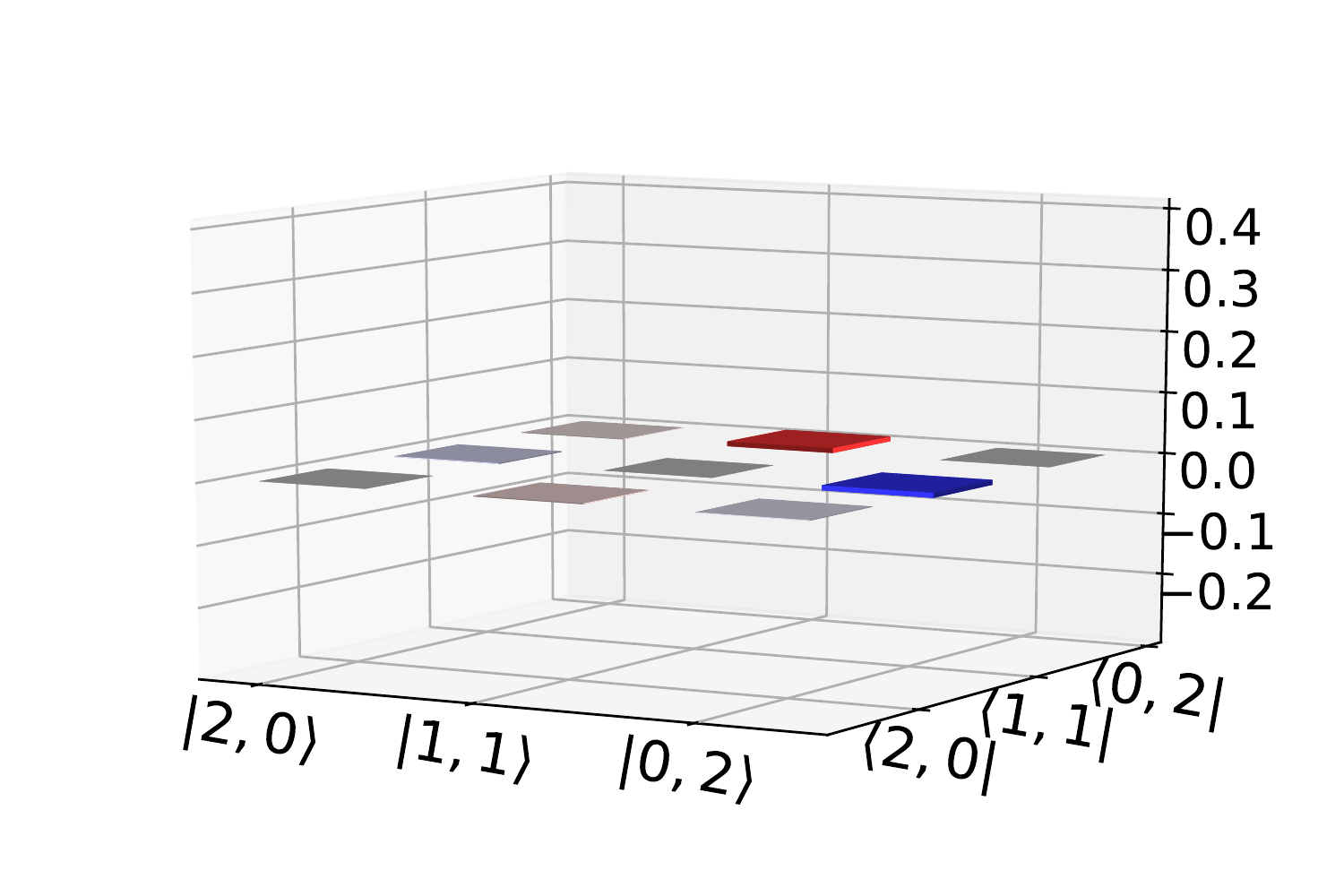}}
\subfigure[Photon number.]{\includegraphics[scale=0.25, clip]{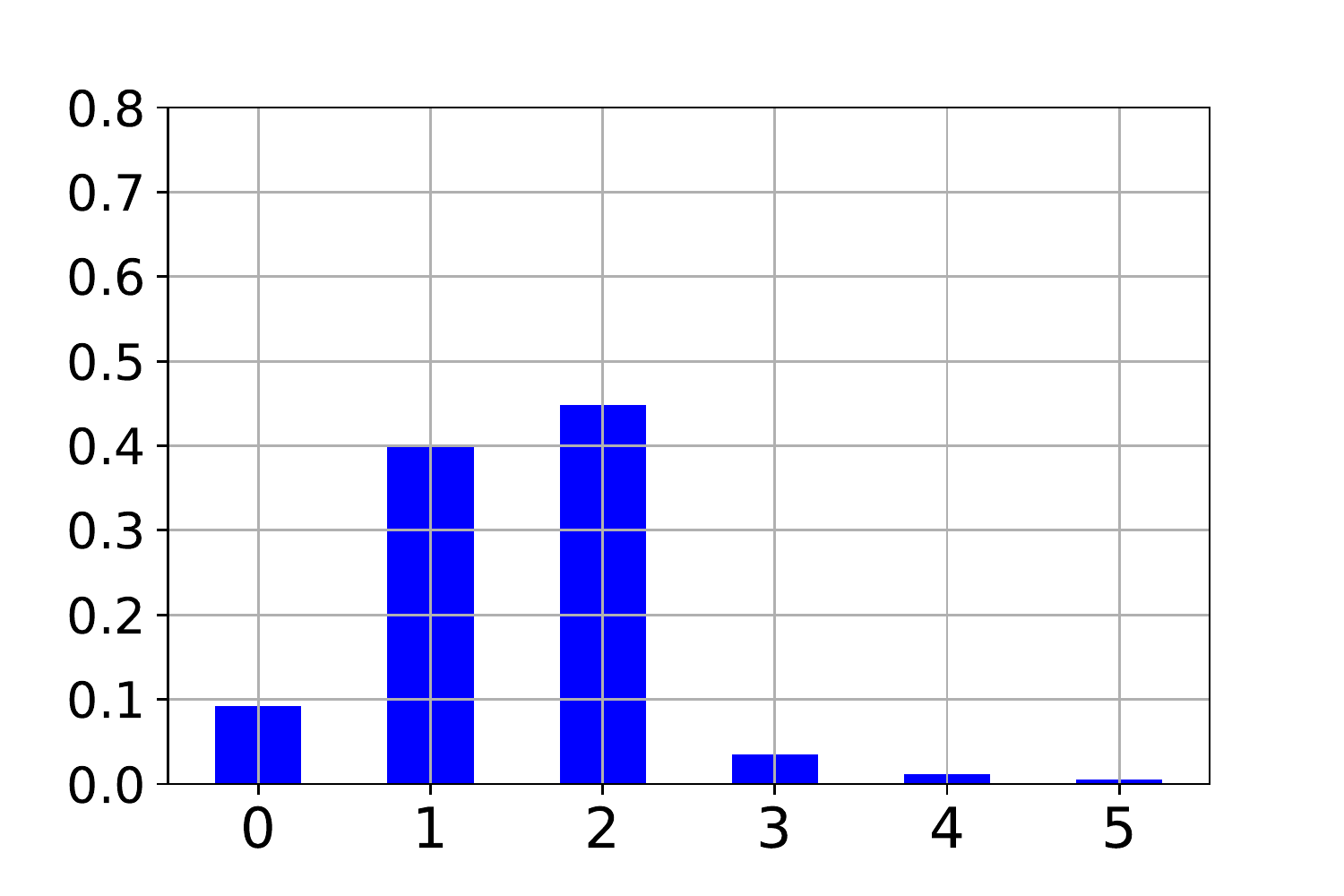}}
\caption{
Experimental density matrix for $(\ket{2,0}+\ket{1,1}+\ket{0,2})/\sqrt{3}$.
The qutrit matrix elements are colored in red (positive) or blue (negative).
}
\label{fig:state1}
\end{figure*}

\begin{figure*}[tb]
\centering
\subfigure[Density matrix (real).]{\includegraphics[scale=0.4, clip]{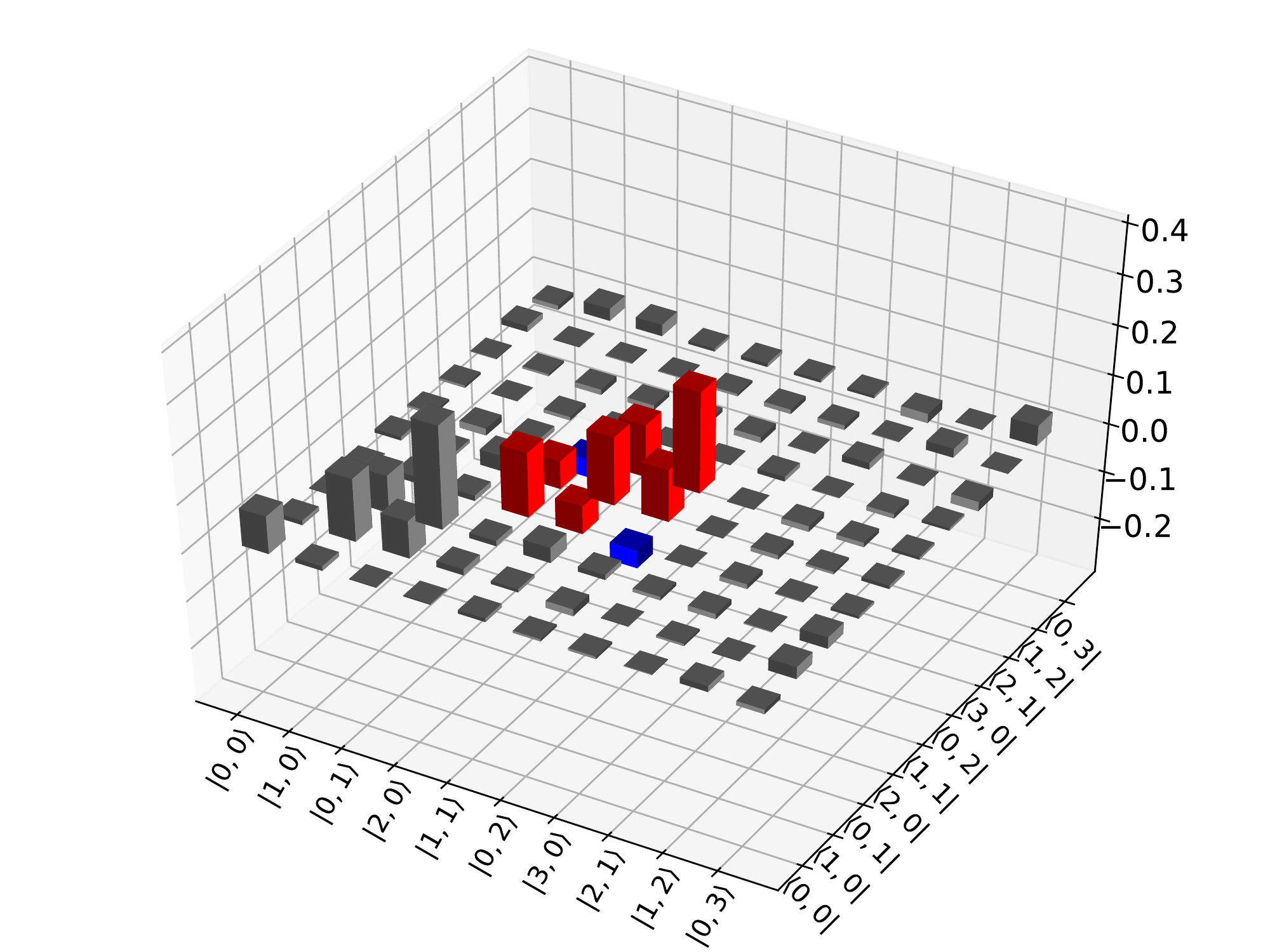}}
\subfigure[Density matrix (imaginary).]{\includegraphics[scale=0.4, clip]{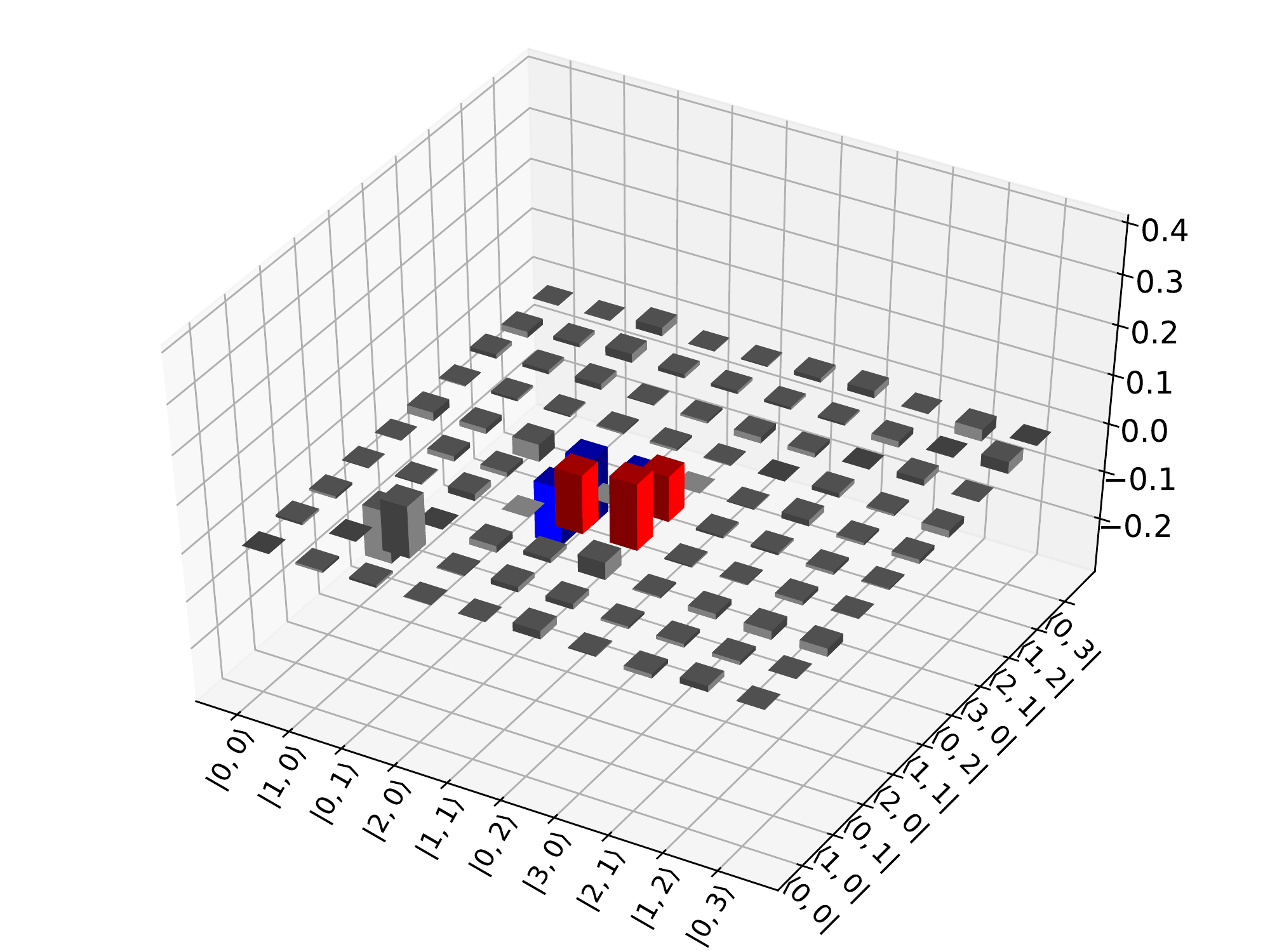}}
\\
\subfigure[Qutrit subspace (real).]{\includegraphics[scale=0.35, clip]{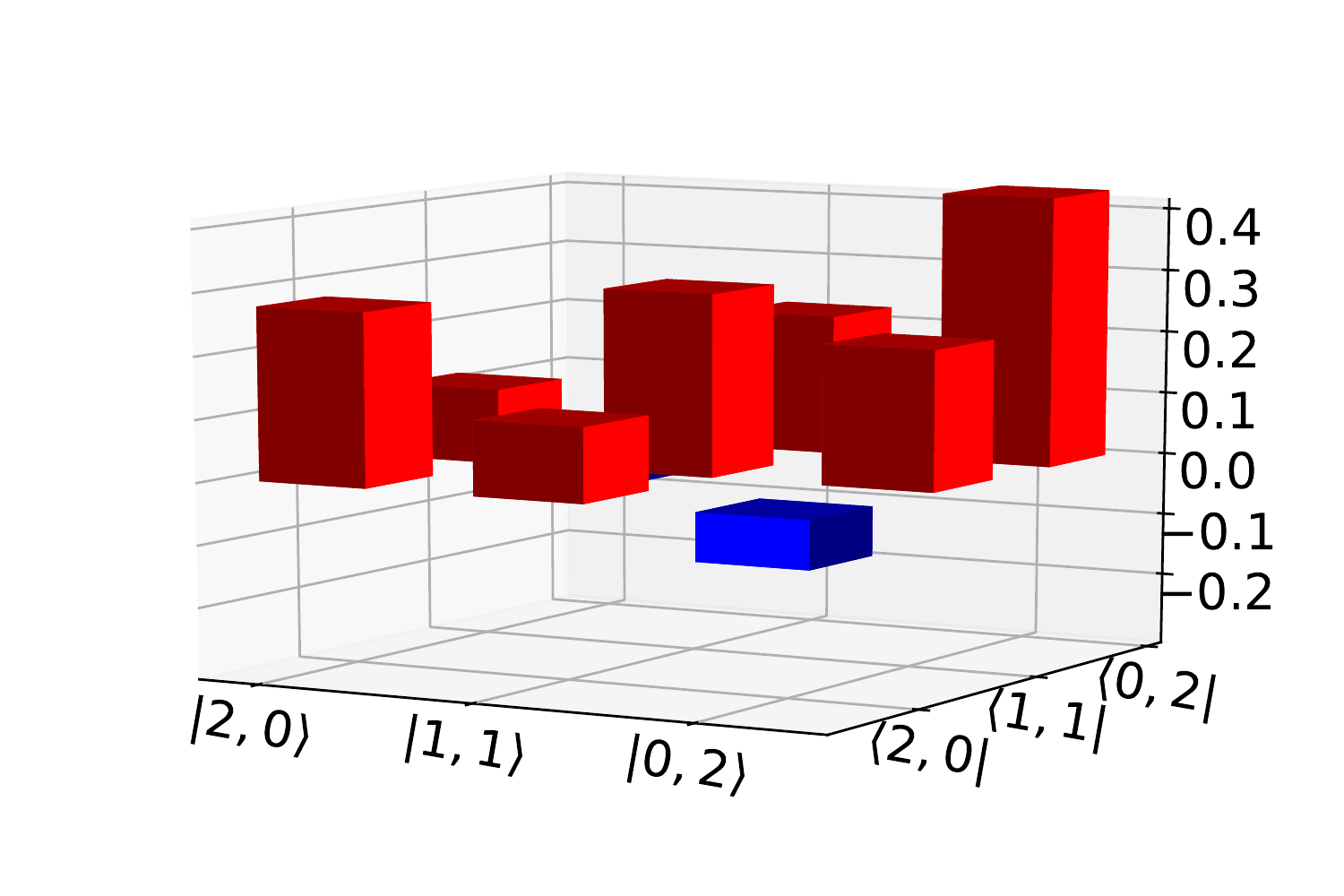}}
\subfigure[Qutrit subspace (imaginary).]{\includegraphics[scale=0.35, clip]{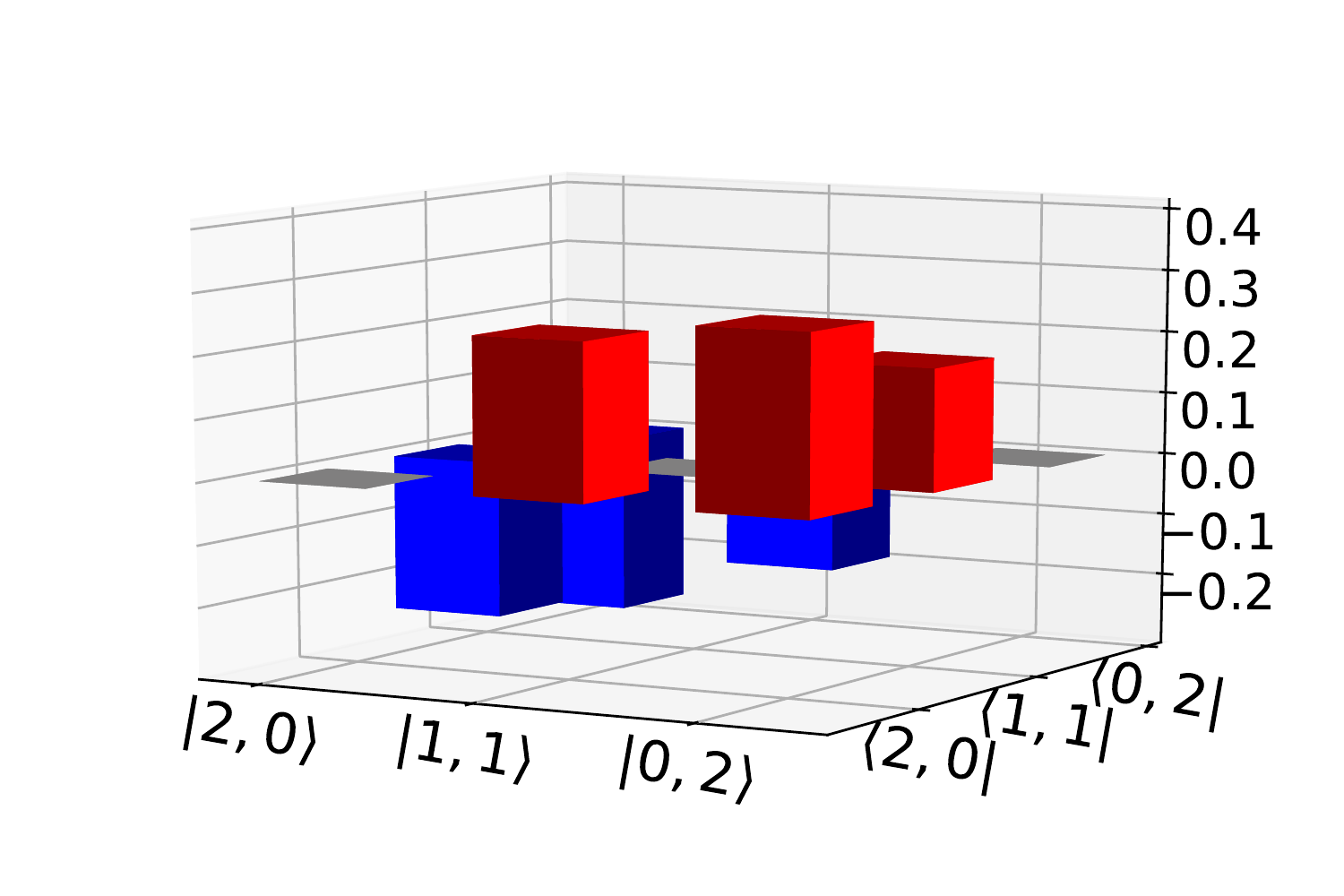}}
\subfigure[Photon number.]{\includegraphics[scale=0.25, clip]{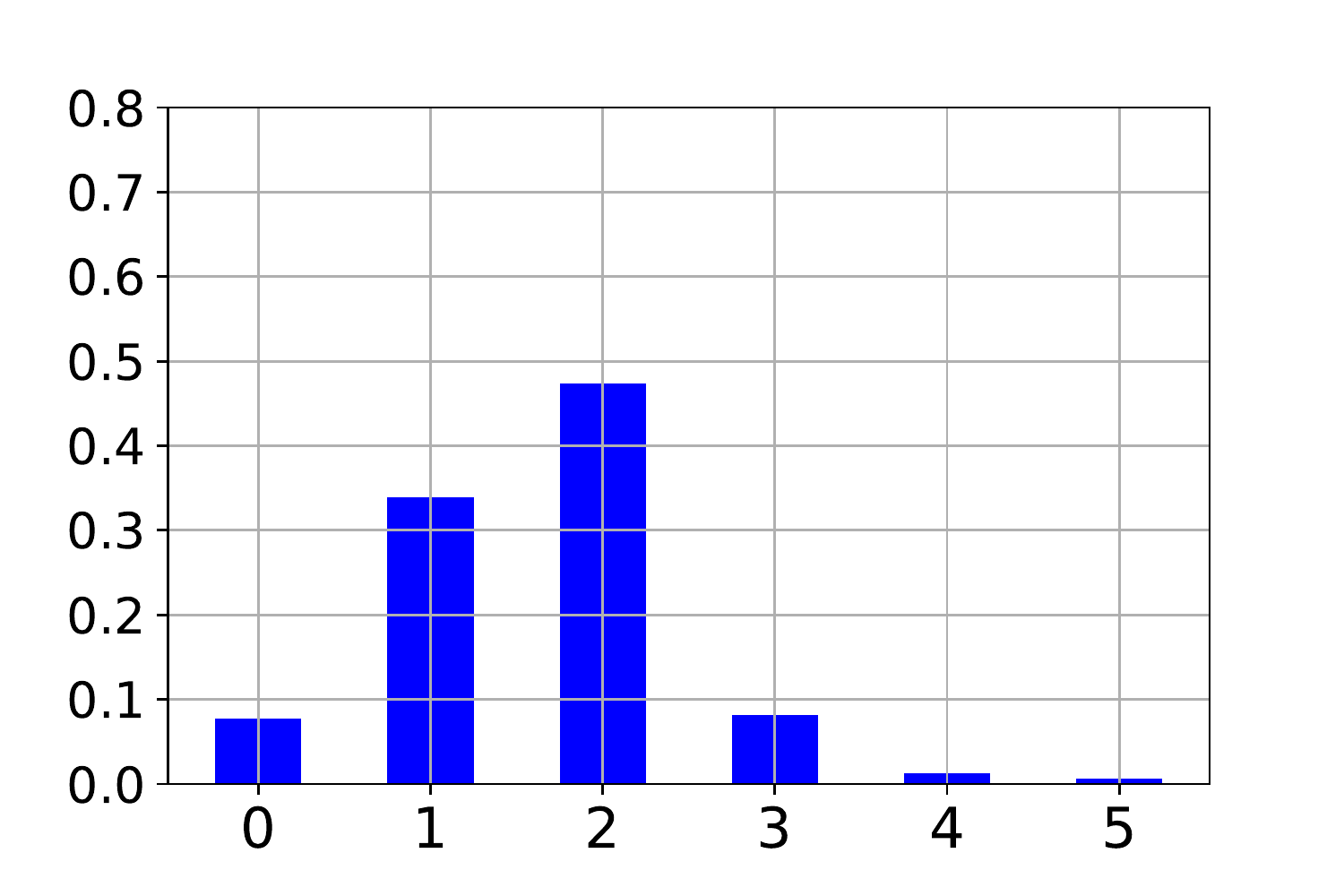}}
\caption{
Experimental density matrix for $[\sqrt{2}\ket{2,0}+(1+\sqrt{2}i)\ket{1,1}+2i\ket{0,2}]/3$. 
The qutrit matrix elements are colored in red (positive) or blue (negative).
}
\label{fig:state2}
\end{figure*}

Note that the code spanned by the two codewords as given in Eq.~\eqref{eq:4photonlosscode} can correct losses of up to one photon. 
In fact, it can be easily seen that the two codewords remain orthogonal after the loss of one
photon in either the first or the second mode. 
Moreover, these two distinct cases lead to logical qubits that live in orthogonal error spaces, $\{\ket{30},\ket{12}\}$ versus $\{\ket{03},\ket{21}\}$, respectively. 
Slightly less simple but also straightforwardly confirmable is that the logical qubit information does not get deformed by the loss of one photon, and so it remains intact in any one of the permitted subspaces. 
However, once two or more photons get lost the supposedly different error spaces start overlapping. 
Thus, the code only works well in the regime of sufficiently small losses. 
More generally, such a two-mode $n$-photon loss code can correct up to $\sqrt{n}-1$ losses, and using our scheme, in principle, any such two-mode code can be experimentally prepared. 
There are also other loss codes that are based upon a higher number of modes where we have seen that our generation scheme may no longer be applicable. 
Nonetheless some class of such multi-mode loss codes makes use of an initial supply of 
so-called NOON states \cite{Bergmann.pra(2016)}. 
For this application, but also for other applications in the context of quantum metrology or lithography, the ability to experimentally prepare NOON states is of great interest. 
We consider this example next.

\subsection{NOON states}

A general NOON state is given by
\begin{equation}
\frac{1}{\sqrt{2}}(\ket{N0}+\ket{0N})=\frac{1}{\sqrt{2}}\left(\frac{a_{1}^{\dagger N}}{\sqrt{N!}}+\frac{a_{2}^{\dagger N}}{\sqrt{N!}}\right)\ket{00}. 
\end{equation}
To be able to apply our scheme for their creation, the polynomial
\begin{equation}
p(x,y) = \frac{1}{\sqrt{2N!}}\left(x^{N}+y^{N}\right) 
\end{equation}
has to be decomposed into linear factors.
The decomposition is given by 
\begin{align}
p(x,y) \coloneqq \frac{1}{\sqrt{2N!}}\prod\limits_{k=0}^{N-1}\left(x-\zeta_{2N}\zeta_{N}^{k}y\right), 
\end{align}
where $\zeta_{N} = \exp(\frac{2\pi i}{N})$ is the $N$-th root of unity.

Therefore, one can write
\begin{align}
p(a_{1}^{\dagger},a_{2}^{\dagger})\ket{00}
&= \frac{1}{\sqrt{2N!}}\prod\limits_{k=0}^{N-1}\left(a_{1}^{\dagger}-\zeta_{2N}\zeta_{N}^{k}a_{2}^{\dagger}\right)\ket{00} \notag\\
&= \frac{1}{\sqrt{2N!}}\sqrt{2}^{N}\prod\limits_{k=0}^{N-1}\left(t_{k}a_{1}^{\dagger}+r_{k}a_{2}^{\dagger}\right)\ket{00}, 
\end{align}
where the corresponding transmission and reflection coefficients are
\begin{align}
t_{k} = \frac{1}{\sqrt{2}} \qquad \text{and} \qquad r_{k} = -\frac{\zeta_{2N}\zeta_{N}^{k}}{\sqrt{2}}. 
\end{align}
The corresponding success probability is 
\begin{align}
p_{N}=q^{2N}(1-q^{2})^{2}\frac{2N!}{2^{N}}\frac{1}{N^{N/2}}.
\end{align}
In Fig. \ref{fig:NOON}, the success probability is shown for various values of $N$. Further examples are presented in the Appendix.

\section{Experiment}
\label{sec:experiment}

Based on the above theory, now we also present an experiment on the generation of two-mode qutrit states ($n=2$, $m=2$). 
The experimental setup is shown in Fig.~\ref{fig:experiment}. 
This is a natural extension of a previous time-bin qubit experiment ($n=1$, $m=2$)~\cite{Takeda.pra(2013)}. 
Instead of preparing two independent nondegenerate optical parametric oscillators (NOPOs), we use a single NOPO, combined with two Mach-Zehnder interferometers on the idler side, with asymmetric arm lengths. 
The NOPO is pumped continuously in a weak-pump regime, generating two-mode squeezed vacuum beams (signal and idler beams) continuously, which have finite correlation times determined by the bandwidth of the NOPO cavity (about 12~MHz of full width at half maximum). 
The signal and idler fields are two frequency modes of the NOPO cavity, separated by a free spectrum range (about 600~MHz).
They are spatially separated by a filter cavity whose round-trip length is half of that of the NOPO. 
Additional two filter cavities further eliminate irrelevant fields before photon detections.
After the filtering of the idler field, there are two Mach-Zehnder interferometers. 
The optical delay lines in the two Mach-Zehnder interferometers are common, implemented with an optical fiber with a length of about 50~m.
Thanks to the delay line, sufficiently time-shifted idler fields interfere before the photon detection, which enables the heralded generation of time-bin superposition states.
When two silicone avalanche photodiodes (APDs) detect a photon simultaneously on the idler side, a qutrit state is heralded on the signal side.
The transmission and reflection coefficients that determine the heralded state (see Appendix~\ref{sec:twophoton}) are controllable by using a half-wave plate (HWP) and two polarization beamsplitters (PBSs) as a variable beamsplitter. 
The angle of the wave plate is adjusted by referring to the photon counting rate from each arm.
We set a time window of about 30~ns to judge two photon detection events to be simultaneous. 
Simultaneous detection events were about 50 times per second. 
Note that the event rate is theoretically state-dependent for the reasons discussed in Sec.~\ref{ssec:bias}, and we surely observed such dependence in the experiment.
The signal field, reflected by the first filter cavity, is sent to the characterization setup. 
The heralded state is fully characterized with quantum tomography employing two homodyne detectors for simultaneous measurements of conjugate quadrature variables \cite{Takeda.pra(2013)}.

Here we show the results of two qutrit states as examples.
One state is $(\ket{2,0}+\ket{1,1}+\ket{0,2})/\sqrt{3}$, and the other one is $[\sqrt{2}\ket{2,0}+(1+\sqrt{2}i)\ket{1,1}+2i\ket{0,2}]/3$. 
The resulting density matrices are shown in Fig.~\ref{fig:state1} and Fig.~\ref{fig:state2}, respectively.
The two-photon component was 45--50\%, which is consistent with the heralded single-photon purity of about 70\%. 
The fidelities regarding the qutrit subspace were 93\% and 95\%, respectively. 
These fidelities will be improved if the precision and stability of our experimental setup are enhanced, such as the polarization in the fiber delay line.

\section{Conclusion}

We proposed a scheme to generate arbitrary qudit states in a heralded fashion, distributing $n$ photons ($d=n+1$) in two modes as a superposition state, based on two-mode squeezed states and photon detections. 
We further discussed an extension of our scheme to $m\ge3$ modes, which may sometimes be possible, but not in general. 
Furthermore, we experimentally demonstrated our scheme by generating some exemplary qutrit states. 
States that can be created with our scheme include important states for quantum information applications, such as NOON states with $N\ge3$ and encoded quantum error correction states to suppress photon loss.

\begin{acknowledgments}
This work was partly supported by CREST of JST, JSPS KAKENHI, and APSA, of Japan. 
PvL and MB acknowledge support from Q.com (BMBF).
\end{acknowledgments}

\appendix

\section{Two-photon states}
\label{sec:twophoton}

\begin{figure}[tb]
\centering
\includegraphics[scale=0.3]{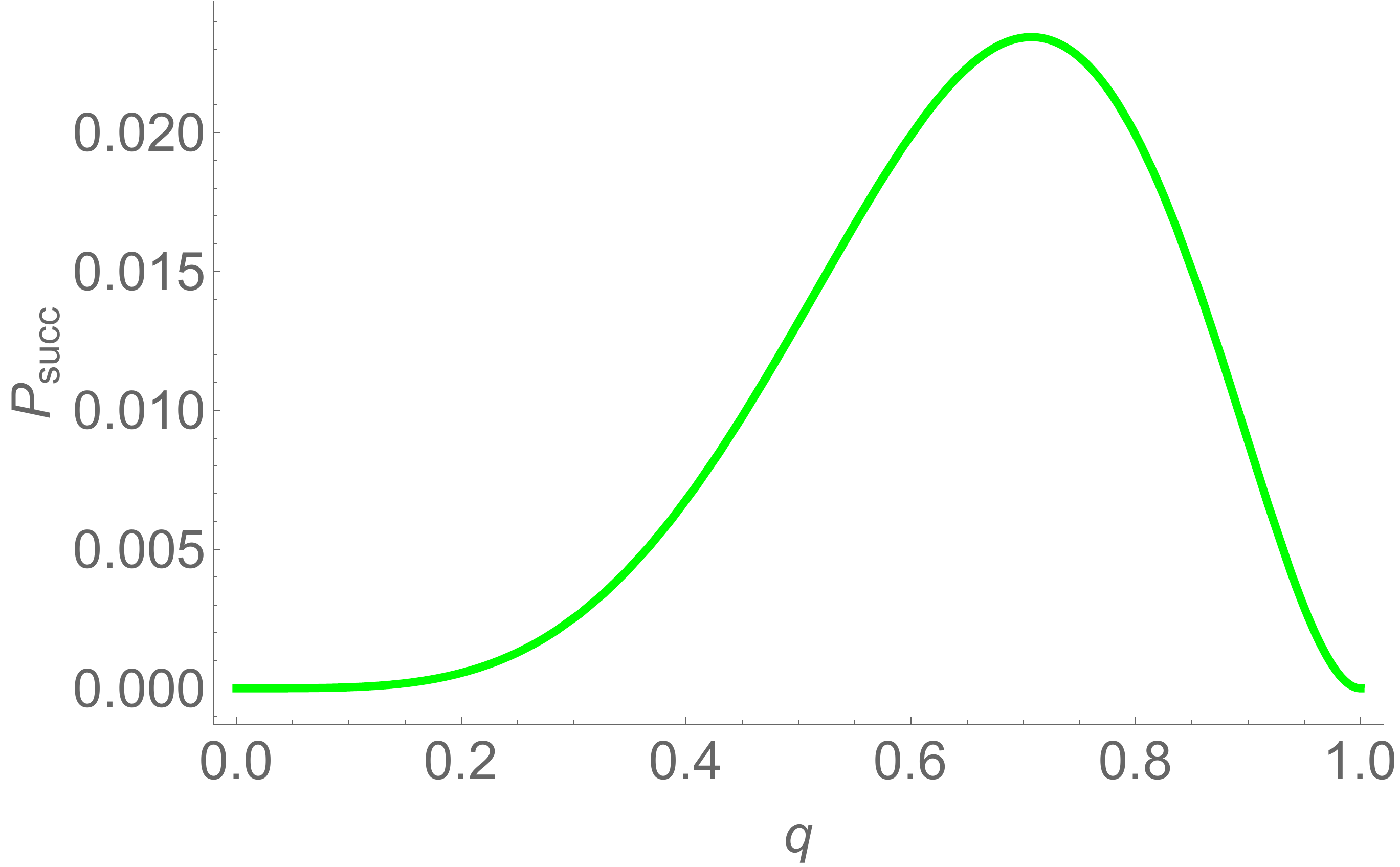}
\caption{
Success probability for creating $\frac{1}{\sqrt{3}}(\ket{20}+\ket{02}+\ket{11})$ in dependence of $q$ (using photon number resolving detectors).
}
\label{fig:N2}
\end{figure}

Let us consider the balanced superposition $\frac{1}{\sqrt{3}}(\ket{20}+\ket{02}+\ket{11})$, which cannot be obtained by linear optics alone. We can write
\begin{align}
& \frac{1}{\sqrt{3}}(\ket{20}+\ket{02}+\ket{11}) \notag\\
& = \frac{1}{\sqrt{3}}\left(\frac{a_{1}^{\dagger 2}}{\sqrt{2}}+\frac{a_{1}^{\dagger 2}}{\sqrt{2}}+a_{1}^{\dagger}a_{2}^{\dagger}\right)\ket{00} \notag\\
& = \sqrt{\frac{1}{6}}\left(a_{1}^{\dagger}+\frac{1}{\sqrt{2}}\left(1-i\right)a_{2}^{\dagger}\right)\left(a_{1}^{\dagger}+\frac{1}{\sqrt{2}}\left(1+i\right)a_{2}^{\dagger}\right)\ket{00} \notag\\
& = \sqrt{\frac{2}{3}}\left(\frac{a_{1}^{\dagger}}{\sqrt{2}}+\frac{1}{2}\left(1-i\right)a_{2}^{\dagger}\right)\left(\frac{a_{1}^{\dagger}}{\sqrt{2}}+\frac{1}{2}\left(1+i\right)a_{2}^{\dagger}\right)\ket{00}.
\end{align}
The probability for successful generation is thus $P_\text{succ}=\frac{3}{8}q^{4}(1-q^{2})^{2}$, which is plotted in Fig.~\ref{fig:N2}.

A general superposition of two-photon states can be decomposed as follows:
\begin{align}
& \alpha\ket{20}+\beta\ket{02}+\gamma\ket{11} \notag\\
= & (\frac{\alpha}{\sqrt{2}}a_{1}^{\dagger2}+\frac{\beta}{\sqrt{2}} a_{2}^{\dagger2}+\gamma a_{1}^{\dagger}a_{2}^{\dagger})\ket{00} \notag\\
= & \frac{\alpha}{\sqrt{2}} \left(a_{1}^{\dagger}-a_{2}^{\dagger}\left(-\frac{\gamma}{\sqrt{2}\alpha}+\sqrt{\frac{\gamma^{2}}{2\alpha^{2}}-\frac{\beta}{\alpha}}\right)\right) \notag\\
& \times \left(a_{1}^{\dagger}-a_{2}^{\dagger}\left(-\frac{\gamma}{\sqrt{2}\alpha}-\sqrt{\frac{\gamma^{2}}{2\alpha^{2}}-\frac{\beta}{\alpha}}\right)\right)\ket{00}.
\end{align}
The corresponding transmission and reflection coefficients are
\begin{subequations}
\begin{align}
t_{1} & = \frac{1}{\sqrt{1+\abs{-\frac{\gamma}{\sqrt{2}\alpha}+\sqrt{\frac{\gamma^{2}}{2\alpha^{2}}-\frac{\beta}{\alpha}}}^{2}}}, \\
r_{1} & = \frac{-\frac{\gamma}{\sqrt{2}\alpha}+\sqrt{\frac{\gamma^{2}}{2\alpha^{2}}-\frac{\beta}{\alpha}}}{\sqrt{1+\abs{-\frac{\gamma}{\sqrt{2}\alpha}+\sqrt{\frac{\gamma^{2}}{2\alpha^{2}}-\frac{\beta}{\alpha}}}^{2}}} \\
t_{2} & = \frac{1}{\sqrt{1+\abs{\frac{\gamma}{\sqrt{2}\alpha}+\sqrt{\frac{\gamma^{2}}{2\alpha^{2}}-\frac{\beta}{\alpha}}}^{2}}}, \\
r_{2} & = -\frac{\frac{\gamma}{\sqrt{2}\alpha}+\sqrt{\frac{\gamma^{2}}{2\alpha^{2}}-\frac{\beta}{\alpha}}}{\sqrt{1+\abs{\frac{\gamma}{\sqrt{2}\alpha}+\sqrt{\frac{\gamma^{2}}{2\alpha^{2}}-\frac{\beta}{\alpha}}}^{2}}}.
\end{align}
\end{subequations}

\section{Three-mode states}

Using the methods described above, as an example of a three-mode state that can be indeed created, we present the following state,
\begin{align}
& \frac{1}{2\sqrt{3}}(a_{1}^{\dagger}+a_{2}^{\dagger})(a_{1}^{\dagger}+a_{3}^{\dagger})(a_{2}^{\dagger}-a_{3}^{\dagger})\ket{000} \notag\\
= & \frac{1}{2\sqrt{3}}(a_{1}^{\dagger 2}a_{2}^{\dagger}-a_{1}^{\dagger 2}a_{3}^{\dagger}-a_{1}^{\dagger 2}a_{3}^{\dagger} + a_{2}^{\dagger 2}a_{1}^{\dagger}+a_{2}^{\dagger 2}a_{3}^{\dagger}-a_{3}^{\dagger 2}a_{2}^{\dagger})\ket{000} \notag\\
= & \frac{1}{\sqrt{3}}\left(\ket{2} \frac{\ket{10}-\ket{01}}{\sqrt{2}}+\ket{1}\frac{\ket{20}-\ket{02}}{\sqrt{2}} + \ket{0} \frac{\ket{21}-\ket{12}}{\sqrt{2}}\right).
\end{align}


\begin{thebibliography}{99}


\bibitem{Yurke.prl(1986)}
B.\ Yurke and D.\ Stoler,
\articletitle{Generating quantum mechanical superpositions of macroscopically distinguishable states via amplitude dispersion,}
\prl 57, 13 (1986).


\bibitem{Hudson.repmathphys(1974)}
R.\ L.\ Hudson,
\articletitle{When is the Wigner quasi-probability density non-negative?}
\repmathphys \textbf{6}, 249 (1974).


\bibitem{Lvovsky.prl(2001)}
A.\ I.\ Lvovsky \etal
\articletitle{Quantum State Reconstruction of the Single-Photon Fock State,}
\prl \textbf{87}, 050402 (2001).


\bibitem{Lvovsky.prl(2002)}
A.\ I.\ Lvovsky and J. Mlynek, 
\articletitle{Quantum-optical catalysis: generating nonclassical states of light by means of linear optics,}
\prl 88, 250401 (2002).


\bibitem{Bimbard.nphoton(2010)}
E.\ Bimbard, N.\ Jain, A.\ MacRae, and A.\ I.\ Lvovsky,
\articletitle{Quantum-optical state engineering up to the two-photon level,}
\natphoton {\bf 4}, 243 (2010).


\bibitem{Yukawa.oe(2013)}
M.\ Yukawa, K.\ Miyata, T.\ Mizuta, H.\ Yonezawa, P.\ Marek, R.\ Filip, and A.\ Furusawa,
\articletitle{Generating superposition of up-to three photons for continuous variable quantum information processing,}
\optexp {\bf 21}, 5529 (2013).


\bibitem{Yukawa.pra(2013)}
M.\ Yukawa, K.\ Miyata, H.\ Yonezawa, P.\ Marek, R.\ Filip, and A.\ Furusawa,
\articletitle{Emulating quantum cubic nonlinearity,}
\pra {\bf 88}, 053816 (2013).


\bibitem{Zavatta.prl(2006)}
A.\ Zavatta, M.\ D'Angelo, V.\ Parigi, and M.\ Bellini,
\articletitle{Remote preparation of arbitrary time-encoded single-photon ebits,}
\prl {\bf 96}, 020502 (2006).


\bibitem{Takeda.pra(2013)}
S.\ Takeda, T.\ Mizuta, M.\ Fuwa, J.\ Yoshikawa, H.\ Yonezawa, and A.\ Furusawa,
\articletitle{Generation and eight-port homodyne characterization of time-bin qubits for continuous-variable quantum information processing,}
\pra {\bf 87}, 043803 (2013).


\bibitem{Ourjoumtsev.science(2006)}
A.\ Ourjoumtsev, R.\ Tualle-Brouri, J.\ Laurat, and P.\ Grangier,
\articletitle{Generating optical Schr\"{o}dinger kittens for quantum information processing,}
\science {\bf 312}, 83 (2006).


\bibitem{Neergaard-Nielsen.prl(2006)}
J.\ S.\ Neergaard-Nielsen, B.\ Melholt Nielsen, C.\ Hettich, K.\ M\/{o}lmer, and E.\ S.\ Polzik,
\articletitle{Generation of a superposition of odd photon number states for quantum information networks,}
\prl {97}, 083604 (2006).


\bibitem{Wakui.oe(2007)}
K.\ Wakui, H.\ Takahashi, A.\ Furusawa, and M.\ Sasaki,
\articletitle{Photon subtracted squeezed states generated with periodically poled KTiOPO$_4$,}
\optexp {\bf 15}, 3568 (2007).


\bibitem{Guerlin.nature(2007)}
C.\ Guerlin, J.\ Bernu, S.\ Del\'eglise, C.\ Sayrin, S.\ Gleyzes, S.\ Kuhr, M.\ Brune, J.-M.\ Raimond, and S.\ Haroche,
\articletitle{Progressive field-state collapse and quantum non-demolition photon counting,}
\nature \textbf{448}, 889 (2007).


\bibitem{Chuang.pra(1997)}
I.\ L.\ Chuang, D.\ W.\ Leung, and Y.\ Yamamoto,
\articletitle{Bosonic quantum codes for amplitude damping,}
\pra 56, 1114 (1997).


\bibitem{Yoshikawa.prx(2013)}
J.\ Yoshikawa, K.\ Makino, S.\ Kurata, P.\ van Loock, and A.\ Furusawa,
\articletitle{Creation, storage, and on-demand release of optical quantum states with a negative Wigner function,}
\prx {\bf 3}, 041028 (2013).


\bibitem{Bimbard.prl(2014)}
E.\ Bimbard \etal
\articletitle{Homodyne tomography of a single photon retrieved on demand from a cavity-enhanced cold atom memory,}
\prl \textbf{112}, 033601 (2014).


\bibitem{Kim.prl(2000)}
Y.-H.\ Kim, R.\ Yu, S.\ P.\ Kulik, Y.\ Shih, and M.\ O.\ Scully,
\articletitle{Delayed ``choice'' quantum eraser,}
\prl {\bf 84}, 1 (2000).


\bibitem{DLCZ}
L.\ M.\ Duan, M.\ D.\ Lukin, J.\ I.\ Cirac, and P.\ Zoller,
\articletitle{Long-distance quantum communication with atomic ensembles and linear optics,}
\nature \textbf{414}, 413 (2001).


\bibitem{Bergmann.pra(2016)}
M.\ Bergmann and P.\ van Loock,
\articletitle{Quantum error correction against photon loss using NOON states,}
\pra 94, 012311 (2016).


\end{thebibliography}
\end{document}